\RequirePackage{fix-cm}
\documentclass[twocolumn,epjc3]{svjour3}  
\smartqed  % flush right qed marks, e.g. at end of proof
\RequirePackage{graphicx}
\newcommand{\kt}{k_{\rm T}}
\newcommand{\ncoll}{n_{\rm coll}}
\newcommand{\al}{a_{\rm L}}
\newcommand{\bl}{b_{\rm L}}
\newcommand{\rpa}{R_{\rm pA}}
\newcommand{\pt}{p_{\rm T}}
\newcommand{\rppb}{R_{\rm pPb}}
\newcommand{\qppb}{Q_{\rm pPb}}
\newcommand{\siglq}{\sigma_{\rm LQ}}
\newcommand{\sighq}{\sigma_{\rm HQ}}
\newcommand{\raa}{R_{\rm AA}}
\newcommand{\snn}{\sqrt{s_{\rm NN}}}
\journalname{Eur. Phys. J. }
\begin{document}

\title{ Investigating $D^0$ meson production in $p-$Pb collisions at 5.02 TeV with a multi-phase transport model%\thanksref{t1}
}
%\subtitle{Do you have a subtitle?\\ If so, write it here}

%\titlerunning{Short form of title}        % if too long for running head

\author{Chao Zhang\thanksref{addr1}\thanksref{e1}
        \and
        Liang Zheng\thanksref{addr2}\thanksref{e2}
        \and
        Shusu Shi\thanksref{addr3}\thanksref{e3}
        \and
        Zi-Wei Lin\thanksref{addr4}\thanksref{e4}
%etc.
}

%\thankstext{t1}{Grants or other notes
%about the article that should go on the front page should be
%placed here. General acknowledgments should be placed at the end of the article.
\thankstext{e1}{e-mail: chaoz@whut.edu.cn}
\thankstext{e2}{e-mail: zhengliang@cug.edu.cn}
\thankstext{e3}{e-mail: shiss@mail.ccnu.edu.cn}
\thankstext{e4}{e-mail: linz@ecu.edu}

%\authorrunning{Short form of author list} % if too long for running head

\institute{School of Science, Wuhan University of Technology, Wuhan, 430070, China \label{addr1}
           \and
           School of Mathematics and Physics, China University of Geosciences (Wuhan), Wuhan 430074, China \label{addr2}
           \and
           Institute of Particle Physics and Key Laboratory of Quark\&Lepton Physics (MOE), Central China Normal University, Wuhan 430079, China \label{addr3}
           \and
           Department of Physics, East Carolina University, Greenville, NC 27858, USA \label{addr4}
          % \emph{Present Address:} if needed\label{addr3}
}

\date{Received: date / Accepted: date}
% The correct dates will be entered by the editor

\maketitle

\begin{abstract}
We study the production of $D^0$ meson in $p$+$p$ and $p-$Pb collisions using the improved AMPT model considering both coalescence and independent fragmentation of charm quarks after the Cronin broadening are included. After a detailed discussion of the improvements implemented in the AMPT model for heavy quark production, we show that the modified AMPT model can provide good description of $D^0$ meson spectra in $p-$Pb collisions, the $\qppb$ data at different centrality and $\rppb$  data in both mid- and  forward (backward) rapidities. We also studied the effects of nuclear shadowing and parton cascade on the rapidity dependence of $D^{0}$ meson production and $\rppb$. Our results indicate that having the same strength of the Cronin (i.e $\delta$ value) obtained from the mid-rapidity data leads to a considerable overestimation of the $D^0$ meson spectra and $\rppb$ data at high $\pt$ in the backward rapidity. As a result, the $\delta$ is determined via a $\chi^2$ fitting of the $\rppb$ data across various rapidities. This work lays the foundation for a better understanding of cold-nuclear-matter (CNM) effects in relativistic heavy-ion collisions.

%\keywords{\and Second keyword \and More}
% \PACS{PACS code1 \and PACS code2 \and more}
% \subclass{MSC code1 \and MSC code2 \and more}
\end{abstract}

\section{Introduction}
\label{intro}

In the past two decades, experiments conducted at the Relativistic Heavy Ion Collider (RHIC) and the Large Hadron Collider (LHC) have accumulated a remarkable group of data providing compelling evidence for the existence of a hot and dense matter known as the quark-gluon plasma (QGP)~\cite{Gyulassy:2004zy,STAR:2005gfr,PHENIX:2004vcz}. The primary goal of high-energy heavy-ion physics is to investigate the properties of the QGP. Heavy flavors serve as valuable probes to explore the properties of the QGP created in relativistic heavy-ion collisions~\cite{vanHees:2005wb,Brambilla:2010cs,Andronic:2015wma}. The mass of heavy flavors are larger than the $\Lambda_{QCD}$ or the temperature achieved in the formed QGP. As a result, heavy quarks are primarily produced from hard scatterings before the QGP formation, and they experience almost the full evolution of the system expansion. In addition, the thermal corrections on the heavy quarks are largely suppressed~\cite{Moore:2004tg}, thus enabling them to better retain memory of their interactions with the medium.

The measurement of heavy hadron production in $p$+$p$ collisions helps to examine our understanding of various aspects of QCD~\cite{LHCb:2016ikn,ATLAS:2015igt,ALICE:2019nxm}. It also provides a crucial baseline for understanding the effects of strongly coupled medium in nuclear collisions. In nucleus-nucleus collisions, heavy quarks or hadrons are expected to interact with the QGP or the hadronic medium through elastic or radiative processes, and the effect can be measured by nuclear modification factors $\raa$~\cite{STAR:2014wif,CMS:2017qjw,ALICE:2018lyv,ALICE:2021rxa} and the anisotropic flow of heavy hadrons, such as the elliptic flow $v_2$~\cite{ALICE:2013olq,STAR:2017kkh,ALICE:2017pbx,CMS:2017vhp,ALICE:2020pvw}. For example, substantial suppression in the nuclear modification factor $\raa$ and/or a notably non-zero elliptic flow ($v_2$) for open heavy particles indicates significant interactions of heavy quarks with the deconfined medium. Several theoretical frameworks and phenomenological models have been developed to describe heavy flavor production in high-energy collisions. The fixed flavor number scheme (FFNS)~\cite{Mangano:1991jk} stands as the simplest approach to heavy flavors within pQCD theory. While the NLO calculations are available for this approach, it does not include gluon fragmentation to heavy flavor hadrons. Other implementations have been developed based on the FFNS method. Results from the general-mass variable-flavor-number scheme (GM-VFNS)~\cite{Kniehl:2004fy,Helenius:2018uul} approach as an extension to FFNS generally agree well with both the $p$+$p$ and $p-$A data in a wide rapidity range. Another widely used pQCD method is the fixed order plus next-to-leading logarithms (FONLL)~\cite{Cacciari:2005rk} formalism. This approach can generally provide a reasonable description of the experimental open charm data, however, its central value often tends to under-predict the data. In addition, the models including the Fokker-Planck approach~\cite{Das:2010tj,He:2012df,Cao:2015hia,Lang:2012nqy} and the relativistic Boltzmann transport approach~\cite{Fochler:2010wn,Djordjevic:2013xoa,Xu:2015bbz,Song:2015ykw,Cao:2016gvr} have been developed to simulate the heavy flavor quarks evolution in the QGP.

To investigate the impact of the hot and dense matter formed in relativistic heavy-ion collisions, it is essential to accurately quantify the effects resulting from the presence of nuclei in the initial state. These effects, often referred to as ``cold-nuclear-matter" (CNM) effects~\cite{Salgado:2011wc}, are of significant interest. In the last two decades, experiments at RHIC and LHC have conducted measurements of heavy flavors in small systems such as $d+$Au and $p-$Pb collisions~\cite{STAR:2004ocv,STAR:2021tte,PHENIX:2013txu,ALICE:2014xjz,CMS:2018loe,CMS:2018duw,LHCb:2017yua,ALICE:2019fhe,ALICE:2016cpm,ALICE:2020wla,LHCb:2023kqs,ALICE:2022exq,LHCb:2022dmh,ALICE:2020wfu}, with the aim of quantifying these CNM effects.
Heavy flavor productions are sensitive to several CNM effects, including: (1) nuclear parton distribution functions (nPDFs)\cite{Eskola:2009uj,Hirai:2007sx}, which modify the parton distribution function in nuclei and affect both initial heavy quark production and the final spectra of heavy flavor hadrons, and (2) nuclear broadening (``Cronin effect")~\cite{Cronin:1974zm,Accardi:2002ik,Vitev:2006bi}, caused by multiple scattering of partons inside the nuclei. The yields of heavy quark and quarkonium production are affected by the CNM within many theoretical methods and phenomenological models. For example, the modification of the parton densities in nuclei are important for mid-rapidity quarkonium production~\cite{Vogt:2010aa}, and pQCD results have indicated the need for the Cronin effect to describe experimental data of open heavy flavors at fixed-target energies~\cite{Mangano:1992kq}. Similarly, in the pQCD-based HVQMNR code\cite{Vogt:2018oje,Vogt:2019xmm,Vogt:2021vsc}, Cronin is also required to describe quarkonium $\pt$ distributions and heavy flavor azimuthal distributions from fixed-target to LHC energies. In our recent work, the longstanding $\rpa$ and $v_2$ puzzle of $D^0$ meson in $p-$Pb collisions at LHC is studied with the Cronin implemented in the AMPT model~\cite{Zhang:2022fum,Zhang:2023xkw}. 

The AMPT model~\cite{Lin:2004en,Lin:2021mdn} is a widely used event generator that is designed to simulate the full phase space evolution of the dense matter created in relativistic heavy-ion collisions. It mainly contain four parts: the fluctuation initial conditions, parton cascade, quark coalescence and hadron cascade. Recently we improved the quark coalescence process~\cite{He:2017tla} and updated modern nuclear parton distribution functions~\cite{Zhang:2019utb} to the AMPT model. we also improved the heavy flavor production~\cite{Zheng:2019alz}, and applied local nuclear scaling to two input parameters for achieving a self-consistent dependence on the system size and centrality~\cite{Zhang:2021vvp}. In this work, we use the AMPT model containing those improvements to study the $D^0$ meson production in the $p-$Pb collisions and $\rppb$ at 5.02 TeV at different rapidities. The rest of the paper is organized as follows: in sec.~\ref{sec:method} we discuss the improvements we made to the heavy flavor production in the AMPT model, in sec.~\ref{sec:res} we calculate the $\pt$ spectra of $D^0$ in both $p$+$p$ and $p-$Pb collision, $\qppb$ of $D^0$ in different centrality  and $\rppb$ of $D^0$ in different rapidities. After discussions in sec.~\ref{sec:dis}, finally we summarize in sec.~\ref{sec:summary}.

\section{The heavy flavor production in the AMPT model}

\label{sec:method}
we have improved the heavy flavor production in the AMPT model by incorporating the following improvements. Firstly, we extract the charm quark jet from the HIJING initial condition, without subjecting them to the string-melting mechanism. Secondly, we adopt the transverse momentum broadening, also known as the ``Cronin effect"\ ~\cite{Cronin:1974zm}, to the initial $Q\bar{Q}$ pairs prior to the parton scattering. Moreover, we supplement the quark coalescence with the implementation of an independent fragmentation process as a means of hadronization. In the following sections, we discuss these improvements in detail.
\subsection{Heavy quark from the initial condition}
\label{sec:2}
\begin{figure}[!htb]
\includegraphics[scale=0.42]{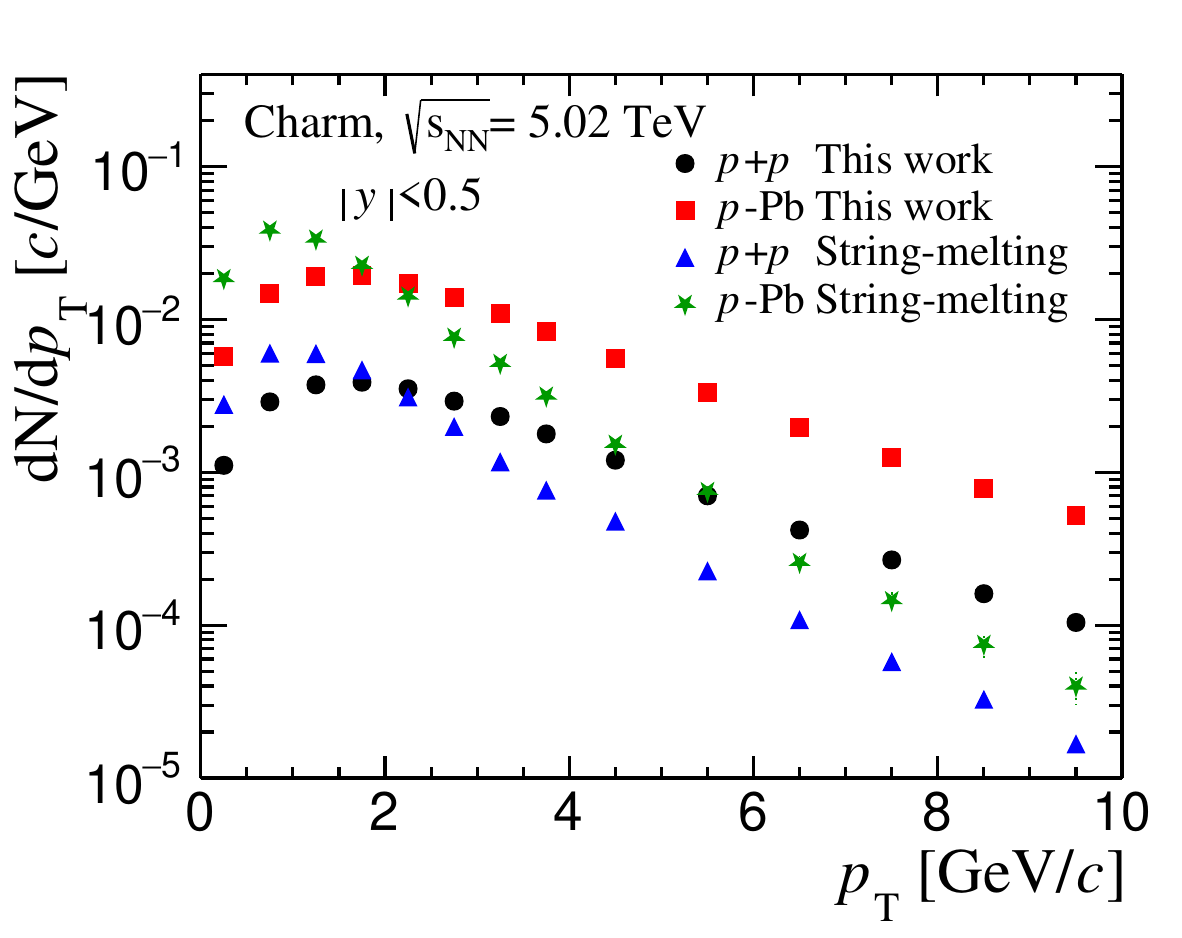}
\caption{ The $p_{T}$ spectra of the $c+\bar{c}$ quarks at mid-rapidity taken from the HIJING initial condition and from string-melting process of the AMPT model in $p$+$p$ and minimum bias $p-$Pb collisions at $\snn=$ 5.02 TeV.
}
\label{fig:1}
\end{figure}

Heavy quarks are believed to be predominantly produced from the primary hard processes because their large mass. Instead of extracting the heavy quarks from the ``melting" of hadrons via string-melting, we take the charm quark from HIJING initial conditions~\cite{Wang:1991hta} and transport them directly to the parton cascade stage to account for their interactions with the de-confined nuclear matter. These initial charm quarks are allowed to interact only after their formation time given by $t_F=E/m_T^2$, where $E$ and $m_T$ are the energy and transverse mass of the heavy quarks, respectively. In Fig.~\ref{fig:1}, We present the initial charm quark transverse momentum ($\pt$) spectra, i.e., prior to their transportation to the parton cascade stage, derived from either this study or the string melting mechanism. These spectra are shown for both $p$+$p$ and $p-$Pb collisions at a center-of-mass energy of 5.02 TeV, aiming to illustrate the impacts of this modification. As depicted, the charm quark transverse momentum spectra taken from initial conditions are much harder compared to those obtained with the string melting in both collision systems.

%%%%%%%%%%%%%%%%%%%%%%%%%%%%%%%%%%%%%%%%%%%%%%%%%%%%%%%%%%%
%%%%%%%%%%%%%%%%%%%%%%%%%%%%%%%%%%%%%%%%%%%%%%%%%%%%%%%%%%%
\subsection{The Cronin effect}
We incorporate the transverse momentum broadening, commonly referred to as the Cronin effect \cite{Cronin:1974zm}, into the initial heavy quarks~\cite{Mangano:1992kq,Vogt:2018oje}. The Cronin effect is frequently considered as the broadening of the transverse momentum of a generated parton resulting from multiple scatterings of incoming partons in the nuclei~\cite{Accardi:2002ik,Kopeliovich:2002yh,Kharzeev:2003wz,Vitev:2006bi}. Hence, its magnitude depends on the number of scatterings endured by a projectile (or target) nucleon during its traversal of the target (or projectile) nucleus.
Therefore in this study, before the heavy quarks undergo the parton scattering process, we include transverse momentum broadening by assigning a transverse momentum kick ($\kt$) to each $Q\bar{Q}$ pair in the initial state. We sample $\kt$ from a two-dimensional Gaussian \cite{Mangano:1992kq,Vogt:2018oje,Vogt:2019xmm,Vogt:2021vsc} with a Gaussian width parameter ($w$):

\begin{eqnarray}
&&f (\vec \kt)=\frac{1}{\pi w^2} e^{-\kt^2/w^2},\label{eq1}\\
&&w=w_0 \sqrt{1+(\ncoll-i) \delta},\label{eq2}
\end{eqnarray}

Here, the value $i = 1$ represents the production of $c\bar{c}$ pairs from a single nucleon participant, while the value $i = 2$ corresponds to the collision between a nucleon participant from the projectile and another from the target. $\ncoll$ denotes the number of primary NN collisions for the participant nucleon in the former case and the total sum of primary NN collisions for both participant nucleons in the latter case. In the case of $p$+$p$ collisions, $w=w_0$. The value of $w_0$ is calculated as shown in Eq. \ref{eq3}~\cite{Zhang:2022fum}.
\begin{equation}
w_0=(0.35{\rm~GeV}/c)~\sqrt {\bl^0(2+\al^0)/\bl/(2+\al)}.\label{eq3}
\end{equation}
In the AMPT model, the original values for the two Lund fragmentation function parameters are $\al^0=0.5$ and $\bl^0=0.9$ GeV$^{-2}$, while the values used in this study are denoted by $\al$ and $\bl$~\cite{Zhang:2021vvp}. The parameter $w_0$, which characterizes the width of Gaussian smearing in the transverse momentum due to the Cronin effect, depends on the Lund parameters. This is because the average squared transverse momentum of a hadron relative to the fragmenting parent string is proportional to the string tension, which scales as $1/\bl/(2+\al)$~\cite{Lin:2004en}. We use $\al=0.8$ and determine $\bl$ according to the local nuclear thickness functions, where the $\bl$ value is $0.7$ GeV$^{-2}$ for $p$+$p$ collisions but smaller for nuclear collisions~\cite{Zhang:2021vvp}. As a result, for $p$+$p$ collisions, $w$ is $0.375$ GeV$/c$, which is close to the original value of $0.35$ GeV$/c$ for the parameter \texttt{PARJ(21)} in the HIJING1.0 model~\cite{Wang:1991hta}. The strength of the Cronin effect is controlled by the parameter $\delta$ in Eq.\ref{eq2}, where in our previous study~\cite{Zhang:2022fum,Zhang:2023xkw} is set to 5.0 in order to describe the mid-rapidity $\rppb$ data of $D^0$ meson. In this work, the same value of $\delta$ is found to overestimate some $\rppb$ data especially at backward rapidity~\cite{LHCb:2017yua}, therefore the $\chi^2$ fit of the $\delta$ is performed using the data at different rapidities. 
%which is set to $\delta=5.0$ for the mid-rapidity and 3.0 for the forward and backward rapidities, based on a fit to the $\rpa$ data of $D^0$ mesons within different rapidities.

To implement the Cronin effect, The $c\bar{c}$ pair is boosted to its rest frame from the nucleon nucleon center-of-mass frame, we then apply a transverse momentum kick to each $c\bar{c}$ pair, sampled from the distribution in Eq. (\ref{eq1}), resulting in an increase in their mean transverse momenta. Finally, the pair is boosted back. Note that this implementation can introduce an artificial peak in the mid-rapidity region of the heavy quark rapidity distribution~\cite{Vogt:2018oje,Vogt:2019xmm,Vogt:2021vsc} because $y={\rm arcsinh}(p_{\rm z}/m_{\rm T})$ will move towards zero as $\pt$ increases. To avoid this, we keep the rapidity of $c\bar{c}$ pairs the same by providing the necessary longitudinal boost after the transverse momentum broadening. We also ensure momentum conservation of the whole parton system in each event by letting the light (anti)quarks share the opposite value of the total $\vec \kt$ given to all $c\bar{c}$ pairs. 

\begin{figure}[!htb]
\includegraphics[scale=0.42]{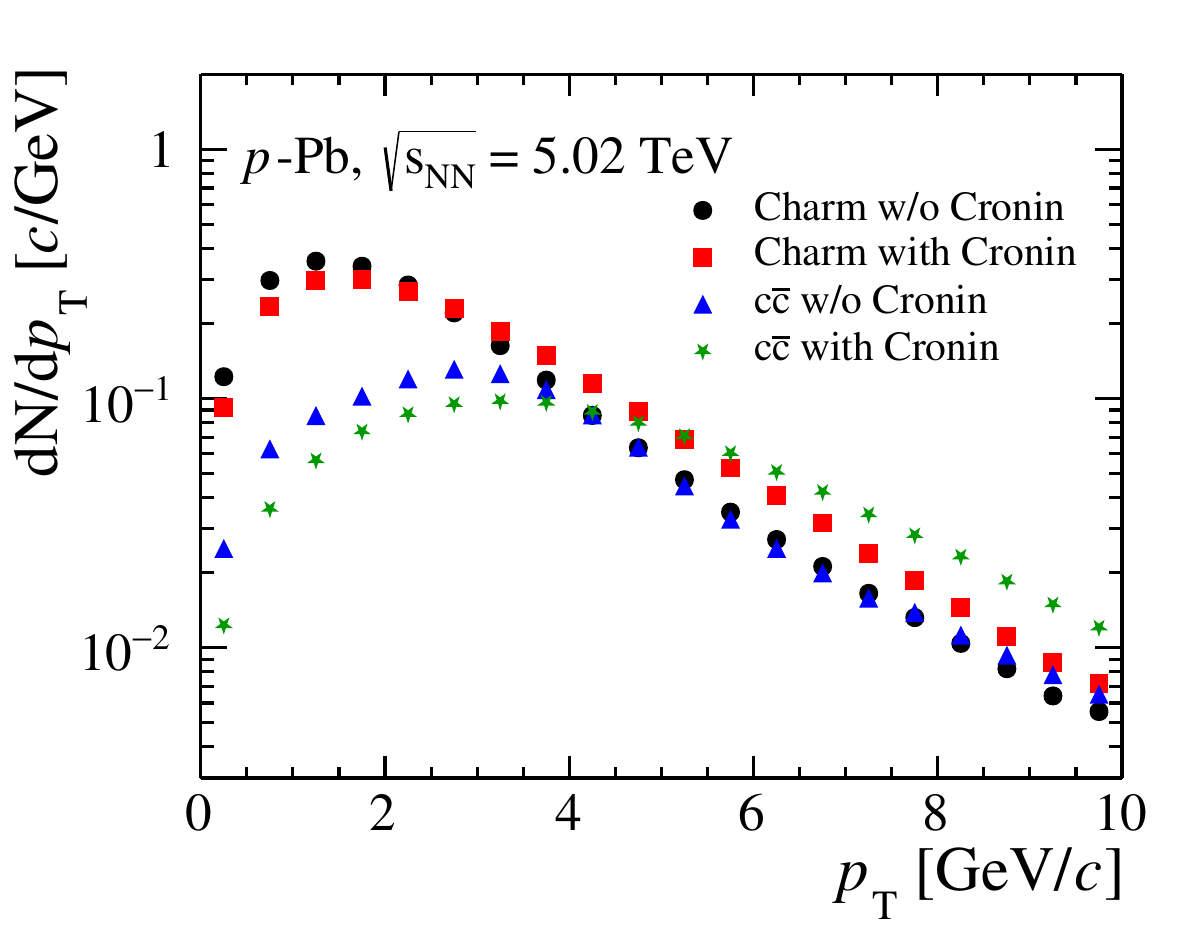}
\caption {The $\pt$ spectra of charm quarks and the total $\pt$ of $c\bar{c}$ pairs prior to the parton cascade from the AMPT model, both with and without the Cronin effect, in minimum bias $p-$Pb collisions at $\snn=$ 5.02 TeV.
}
\label{fig:2}
\end{figure}

Figure~\ref{fig:2} shows the $\pt$ spectra of single charm quarks and the total $\pt$ (\vec{p_T^c}+\vec{p_T^{\bar{c}}}) of $c\bar{c}$ quark pairs in $p-$Pb collisions at 5.02 TeV, obtained from the AMPT model with and without the Cronin broadening. It is evident that the Cronin effect leads to an increase in charm quark $\pt$, and this effect is more pronounced for the charm quark pairs. It is intriguing to observe that the $\pt$ spectra of $c\bar{c}$ pairs without Cronin broadening (blue triangle) achieve a magnitude similar to that of the individual charm quarks (black circle) at high $\pt$. This phenomenon arises from the asymmetrical shape of the $c\bar{c}$ pairs produced from the parton shower, while the $c\bar{c}$ component produced from the perturbative process exhibits a back-to-back configuration.

%%%%%%%%%%%%%%%%%%%%%%%%%%%%%%%%%%%%%%%%%%%%%%%%%%%%%%%%%%%
%%%%%%%%%%%%%%%%%%%%%%%%%%%%%%%%%%%%%%%%%%%%%%%%%%%%%%%%%%%

\subsection{Hadronization}

In high-energy heavy-ion collisions, it is more probable for the low-$\pt$ heavy quarks to combine with thermal partons from the QGP to form new hadrons~\cite{Cao:2016gvr}, this is known as the parton coalescence process. On the other hand, high-$\pt$ heavy quarks are more likely to fragment into low-energy partons, which then form the hadronic bound state of a heavy hadron. This process is referred as the fragmentation process. The above-mentioned two mechanisms will jointly complete the hadronization process of heavy quarks. 
%%%%%%%%%%%%%%%%%%%%%%%%%%%%%%%%%%%%%%%%%%%%%%%%%%%%%%%%%%%
\begin{figure}[!htb]
\includegraphics[scale=0.42]{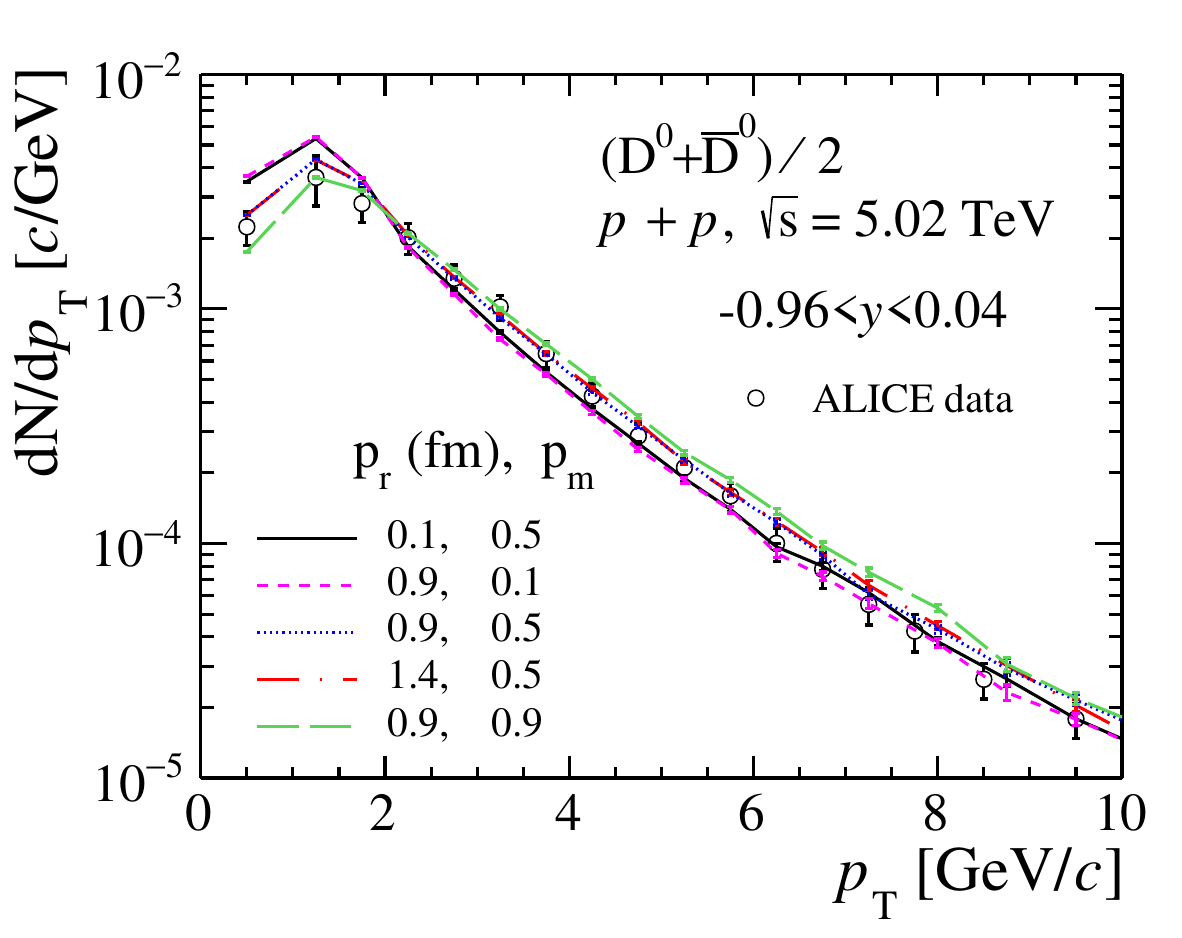}
\caption {The comparison of the $\pt$ spectra of $D^0$ mesons generated using various criteria in the ``two-component" process from AMPT in comparison with the ALICE data~\cite{ALICE:2019nxm,ALICE:2019fhe} in $p$+$p$ collisions at $\sqrt{s}=$ 5.02 TeV.}
\label{fig:3}
\end{figure}
%%%%%%%%%%%%%%%%%%%%%%%%%%%%%%%%%%%%%%%%%%%%%%%%%%%%%%%%%%%

In this study, we implemented coalescence and fragmentation for heavy quark hadronization in the AMPT model by utilizing the PYTHIA~\cite{Sjostrand:1993yb} independent fragmentation process. Specifically, after the partons stopped interactions in the parton cascade, they prepare to undergo quark coalescence for hadronization. Then the heavy hadrons are selected based on the following criteria for the relative distance and invariant mass of the heavy quarks and their coalescence partners:
\begin{eqnarray}
&&d<p_{r},\label{eq4}\\
&&m_{inv}<\sum m_Q+p_{m}(m_H-\sum m_Q) .\label{eq5}
\end{eqnarray}
%%%%%%%%%%%%%%%%%%%%%%%%%%%%%%%%%%%%%%%%%%%%%%%%%%%%%%%%%%%

\begin{figure*}
    \begin{minipage}[t]{0.5\linewidth}
    \centering
    \includegraphics[scale=0.44]{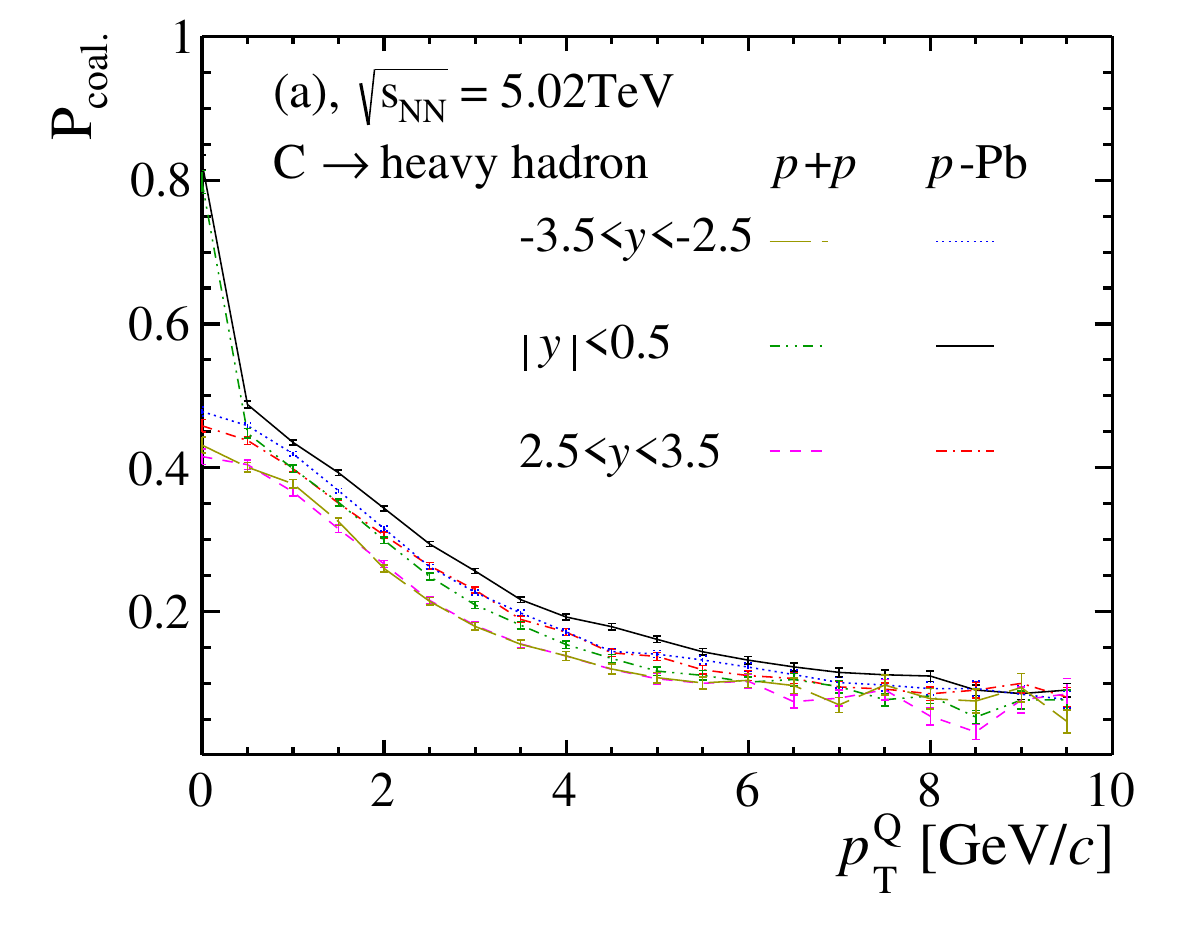}    
  \end{minipage}%
  \begin{minipage}[t]{0.5\linewidth}
    \centering
    \includegraphics[scale=0.44]{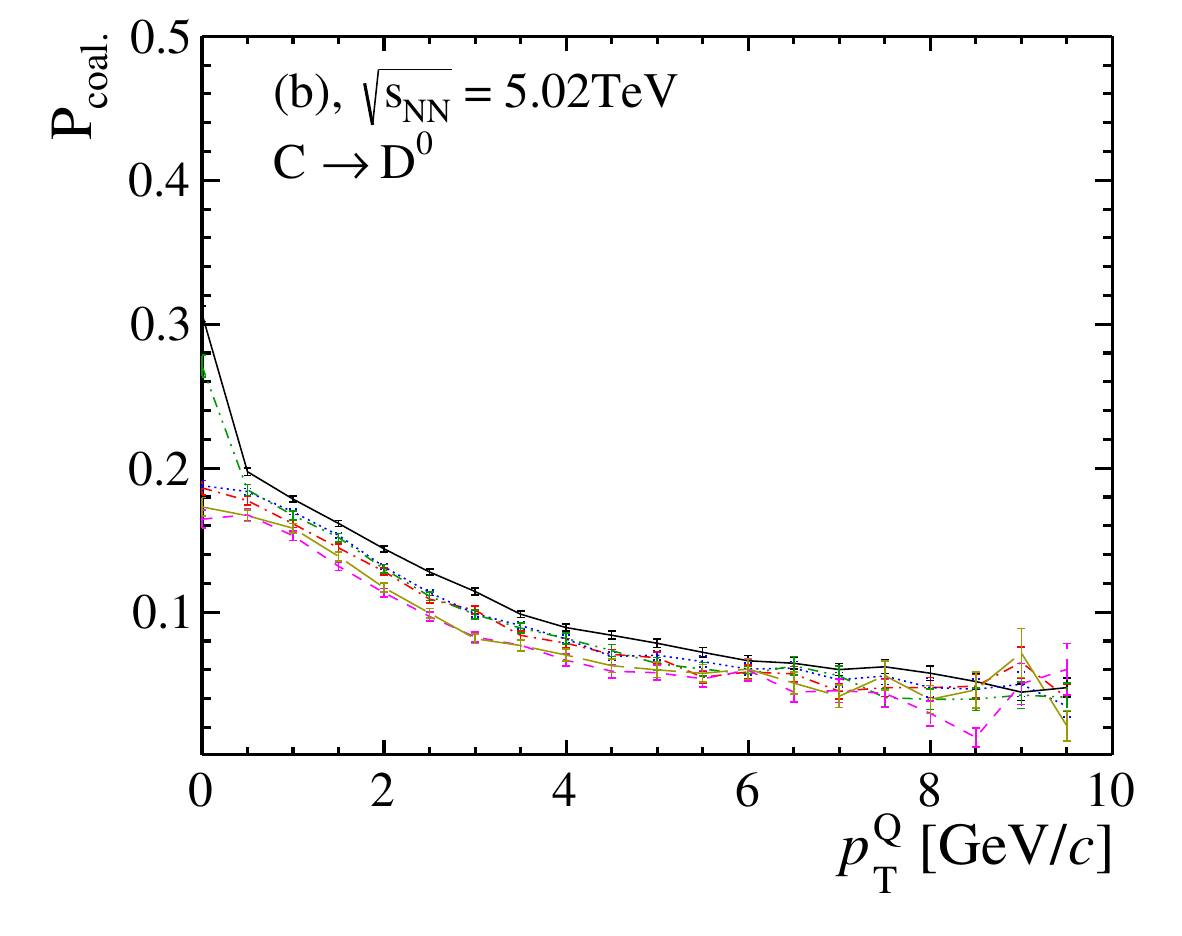}
     \end{minipage}
    \caption{The coalescence probability of charm quarks at different rapidity to (a): charm hadrons and (b): $D^0$ meson, displayed as a function of charm quark transverse momentum in $p$+$p$ and $p-$Pb collisions at $\snn=$5.02 TeV.
}
     \label{fig:4}
  \end{figure*}
%%%%%%%%%%%%%%%%%%%%%%%%%%%%%%%%%%%%%%%%%%%%%%%%%%%%%%%%%%%

Here, $d$ and $m_{inv}$ represent the relative distance and invariant mass, respectively, of the coalescing partners within their rest frame. Furthermore, $m_Q$ and $m_H$ indicates the mass of heavy quarks and hadrons in the pre-coalescence. The parameters $p_{r}=0.9\;fm$ and $p_{m}=0.5$ are determined from a $\chi^2$ fit to the experimental data of $D^0 $ meson $\pt$ spectrum in $p$+$p$ collisions. The heavy quark and its coalescence partner fulfilling both Eq.~\ref{eq4} and~\ref{eq5} are considered suitable for coalescence, while others are subject to independent fragmentation. Fig.~\ref{fig:3} shows the examples of the $D^0 $ meson $\pt$ spectrum using various criteria in the hadronization process. One see that with larger values of $p_{r}$ and $p_{m}$ parameters, the $\pt$ spectrum of $D^0 $ meson becomes harder due to more charm quarks hadronizing via the coalescence process.  

The widely used Peterson fragmentation functions is employed in this study for the heavy quark fragmentation:
\begin{equation}
f(z)\propto\frac{1}{z(1-\frac{1}{z}-\frac{\epsilon_Q}{1-z})^{2}}.\label{eq6}
\end{equation}
where $z$ is the fraction of $E+p_z$ taken by the fragmented hadron out of the parent quark. $\epsilon_Q=0.05$ for the charm and $0.005$ for the bottom.

For the charm hadron candidates obtained in the coalescence process which do not satisfy the requirements in Eq.~\ref{eq4} and~\ref{eq5}, we take the heavy quark $Q$ from the meson candidate $Q\bar{q_1}$, it's important to note that in the case of charmonium, $\bar{q}_1$ could represent $\bar{Q}$, resulting in both pairs of heavy quarks being fragmented separately. A new $q_2\bar{q}_2$ pair is generated out of the vacuum, then the $Q\bar{q_2}$ forms a new meson with its energy and momentum given by the Peterson fragmentation functions as shown above. The flavor of the $q_2\bar{q_2}$ pair is selected with a default probability, $u\bar{u}:d\bar{d}:s\bar{s}=1:1:0.3$~\cite{Sjostrand:1993yb}. The independent fragmentation technique does not produce any additional heavy quarks. The remaining $q_2$ combines with the coalescence partner $\bar{q_1}$ and undergoes the string fragmentation process for hadronization. As for the baryon formation, a diquark $qq$ might be picked instead of the single $q$ by the $Q$, this is known as the ``diquark picture" in PYTHIA~\cite{Sjostrand:1993yb}.

In Fig.~\ref{fig:4}(a) and (b) , we present the coalescence probability as function of charm quark $\pt$ of forming heavy hadrons or $D^0$ mesons, respectively. These results are depicted across three rapidity ranges in both $p$+$p$ and $p-$Pb collisions at 5.02 TeV from the improved AMPT model. It is found that the high $\pt$ charm quarks are more likely to hadronize via the independent fragmentation while the coalescence becomes more dominant in the low $\pt$ region or the larger collisions systems. In addition, the coalescence probability substantially increases for the mid-rapidity charm quarks compared to those in the forward/backward regions, particularly at low $\pt$ ($\leq$0.5 GeV/$c$). In $p-$Pb collisions, we observe higher coalescence probabilities in the backward rapidity region compared to the forward rapidity, whereas in $p$+$p$ collisions, the coalescence probabilities are symmetric for forward and backward rapidity regions. This discrepancy can be attributed to the asymmetric nature of the collision system, which results in variations in multiplicity. Moreover, we observe a similar behavior in coalescence probability between charm quarks to heavy hadrons or to $D^0$ meson, except for the smaller values in the latter case.

%%%%%%%%%%%%%%%%%%%%%%%%%%%%%%%%%%%%%%%%%%%%%%%%%%%%%%%%%%%
\begin{figure}[!htb]
\includegraphics[scale=0.43]{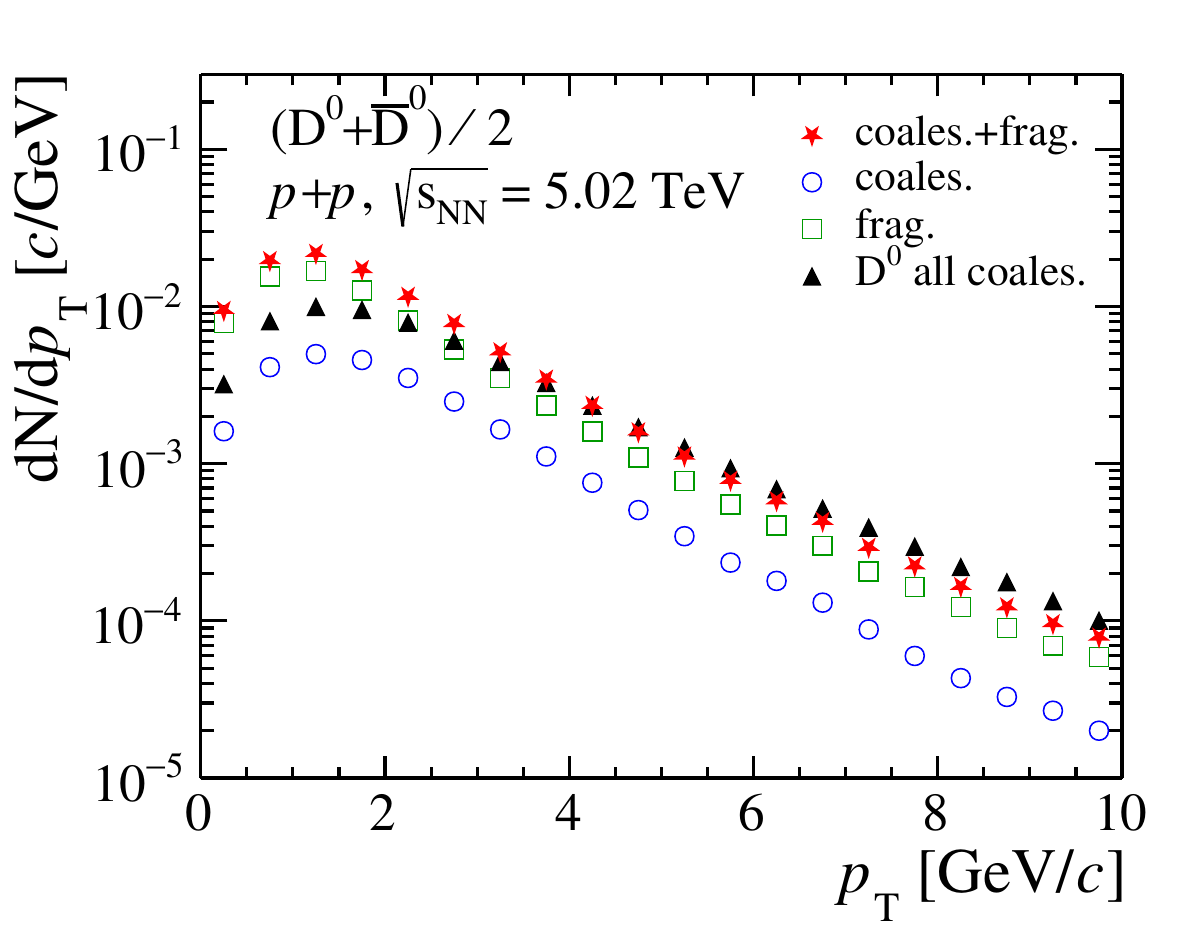}
\caption { The yield of $D^0$ meson generated from different hadronization mechanisms in $p+p$ collisions at $\sqrt{s}$= 5.02 TeV; the black triangle represents the $D^0$ yield from the same final state charm quarks (i.e, after parton cascade) when the independent fragmentation process is switched off.
}
\label{fig:5}
\end{figure}
%%%%%%%%%%%%%%%%%%%%%%%%%%%%%%%%%%%%%%%%%%%%%%%%%%%%%%%%%%%
Figure~\ref{fig:5} shows the yield from coalescence and/or fragmentation to the production of $D^0$ meson in 5.02 TeV $p$+$p$ collisions. It is observed that the production of $D^0$ mesons is predominantly governed by fragmentation at low $\pt$, while the quark coalescence becomes relatively important at intermediate to high transverse momenta. The black triangle represents the $D^0$ mesons formed by the same final state charm quarks if we turn off the independent fragmentation. It is thus natural to see that the hadron spectrum becomes much harder, since the coalescence process tends to increase the charm hadron $\pt$ with respect to its mother heavy quarks, while the charm hadron $\pt$ is usually reduced during the independent fragmentation process.

%%%%%%%%%%%%%%%%%%%%%%%%%%%%%%%%%%%%%%%%%%%%%%%%%%%%%%%%%%%
\begin{figure}[!htb]
\includegraphics[scale=0.42]{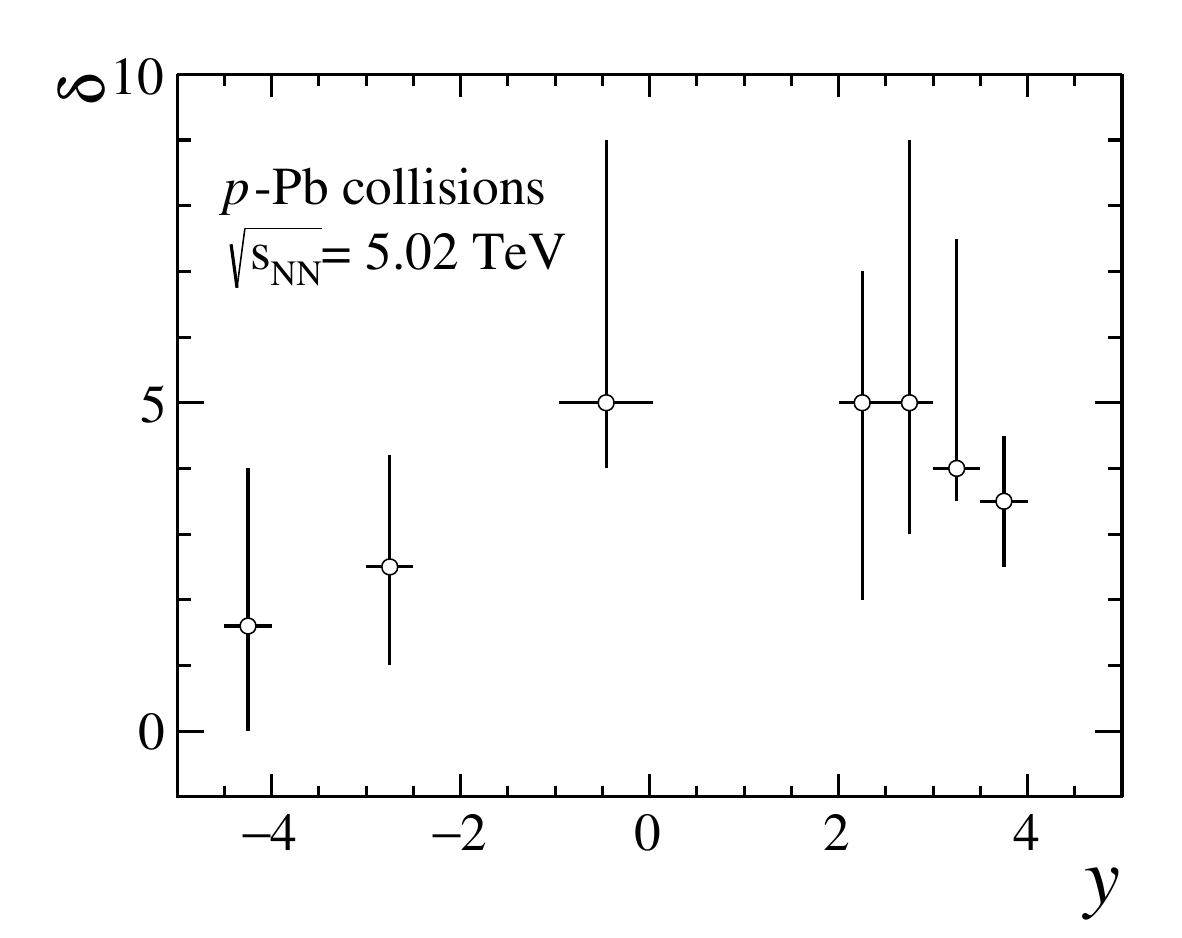}
\caption { The parameter delta in the Cronin extracted from the $\chi^2$ fit of the $\rppb$ data~\cite{LHCb:2016ikn,LHCb:2017yua,ALICE:2019fhe} at different rapidity intervals in $p-$Pb collisions at $\snn=$ 5.02 TeV, where the error bars represent the 95$\%$ confidence interval.}
\label{fig:6}
\end{figure}
%%%%%%%%%%%%%%%%%%%%%%%%%%%%%%%%%%%%%%%%%%%%%%%%%%%%%%%%%%%

In addition to the aforementioned improvement to the heavy flavor production within the AMPT model, we introduce a separate cross section among light quarks ($\siglq$) from that between a heavy quark and other quarks ($\sighq$). This differentiation is prompted by the general difference in scattering cross sections between charm quarks and light (u, d, s) quarks. Default values of  $\siglq$ = 0.5 mb and $\sighq$ = 1.5 mb are adopted unless specified otherwise~\cite{Zhang:2022fum}. These values are established through fitting to charged hadron $v_2$ data in high multiplicity $p-$Pb collisions at 5.02 TeV for $\siglq$, and $D^0$ meson $v_2$ data in high multiplicity $p-$Pb collisions at 8.16 TeV for $\sighq$.

\section{Results}
\label{sec:res}
\begin{figure*}
    \begin{minipage}[t]{0.5\linewidth}
    \centering
    \includegraphics[scale=0.43]{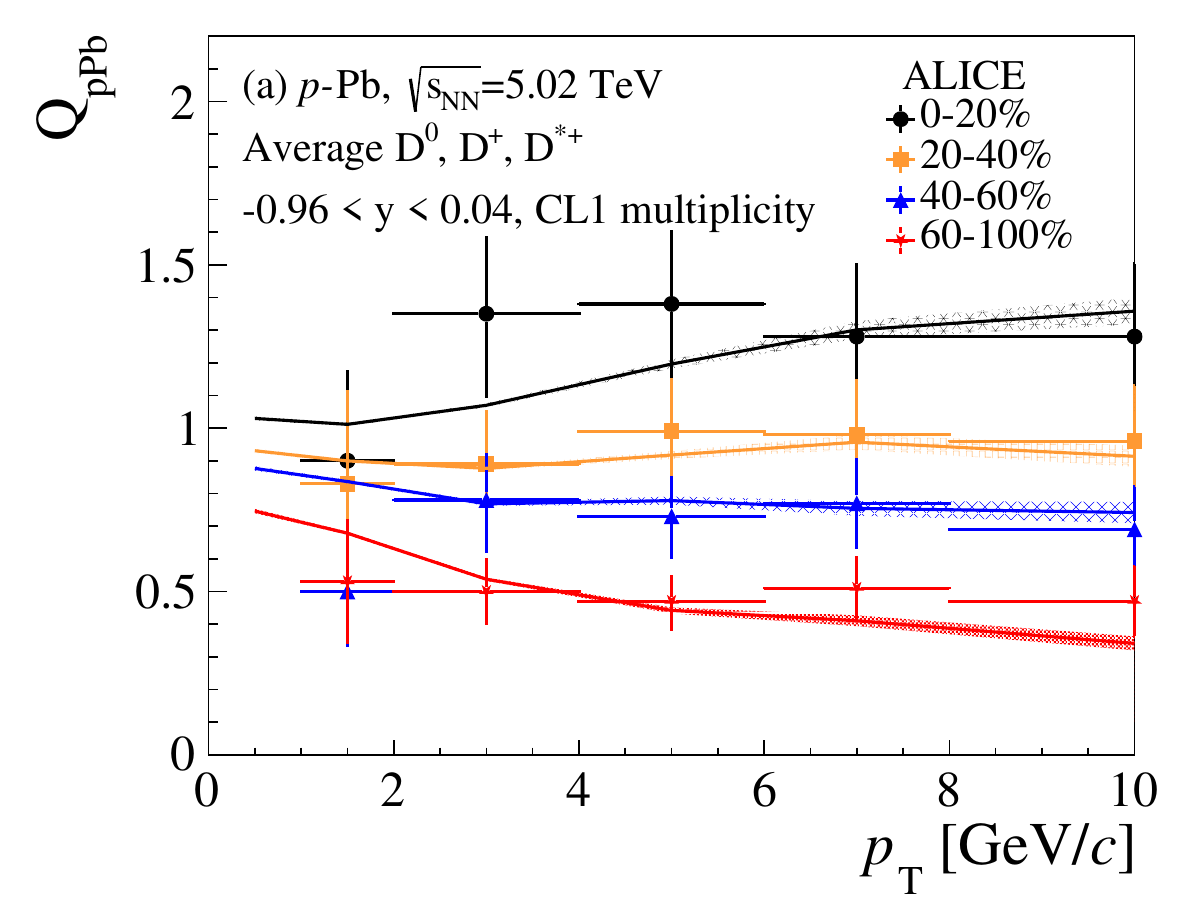}    
  \end{minipage}%
  \begin{minipage}[t]{0.5\linewidth}
    \centering
    \includegraphics[scale=0.43]{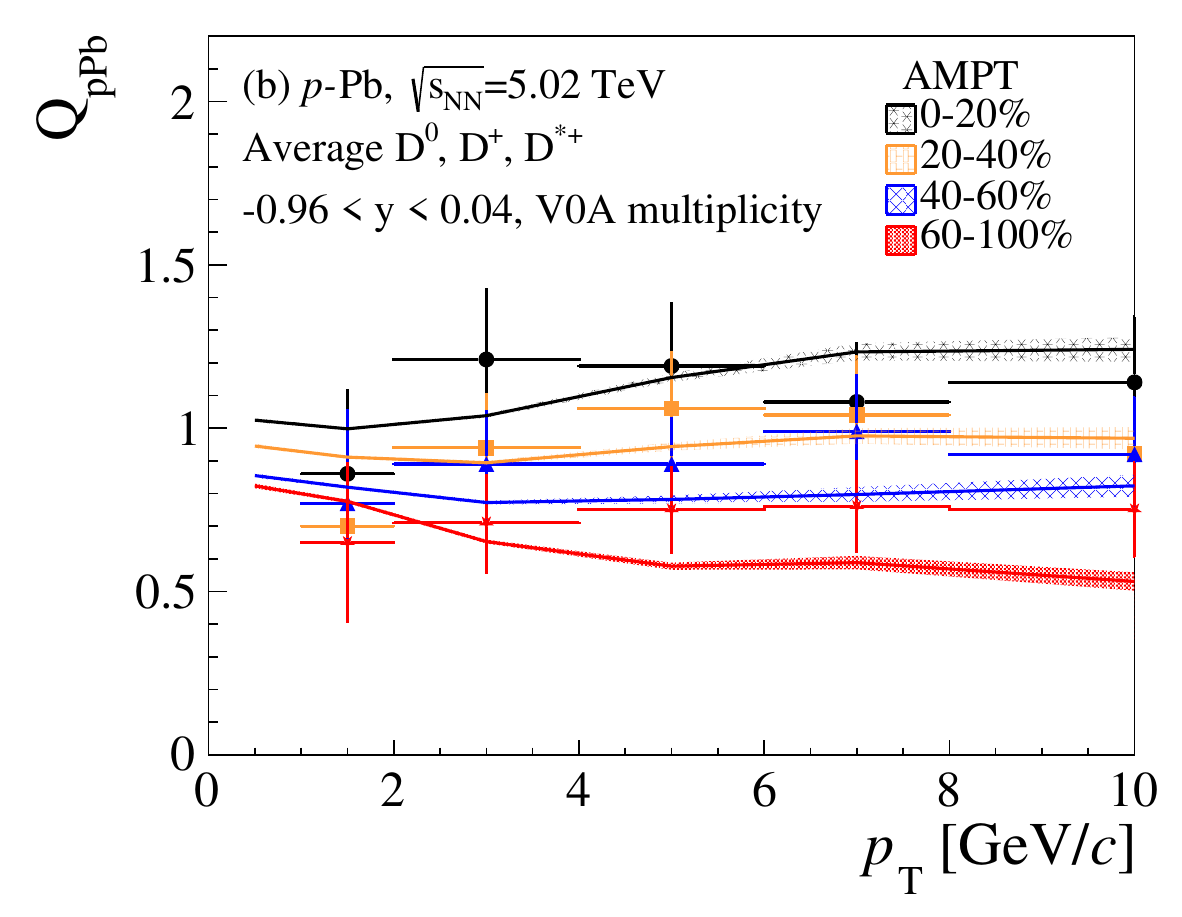}
     \end{minipage}
    \caption{The $\qppb$ from (a): CL1 and (b): V0A event class of average D-mesons as function of the $\pt$ at 0-20, 20-40, 40-60 and 60-100$\%$ centrality intervals from the AMPT model in comparison with the ALICE data~\cite{ALICE:2016cpm} in $p-$Pb collisions at $\snn=$ 5.02 TeV.}
     \label{fig:7}
  \end{figure*}
  
%%%%%%%%%%%%%%%%%%%%%%%%%%%%%%%%%%%%%%%%%%%%%%%%%%%%%%%%%%%

\begin{figure}[!htb]
\includegraphics[scale=0.43]{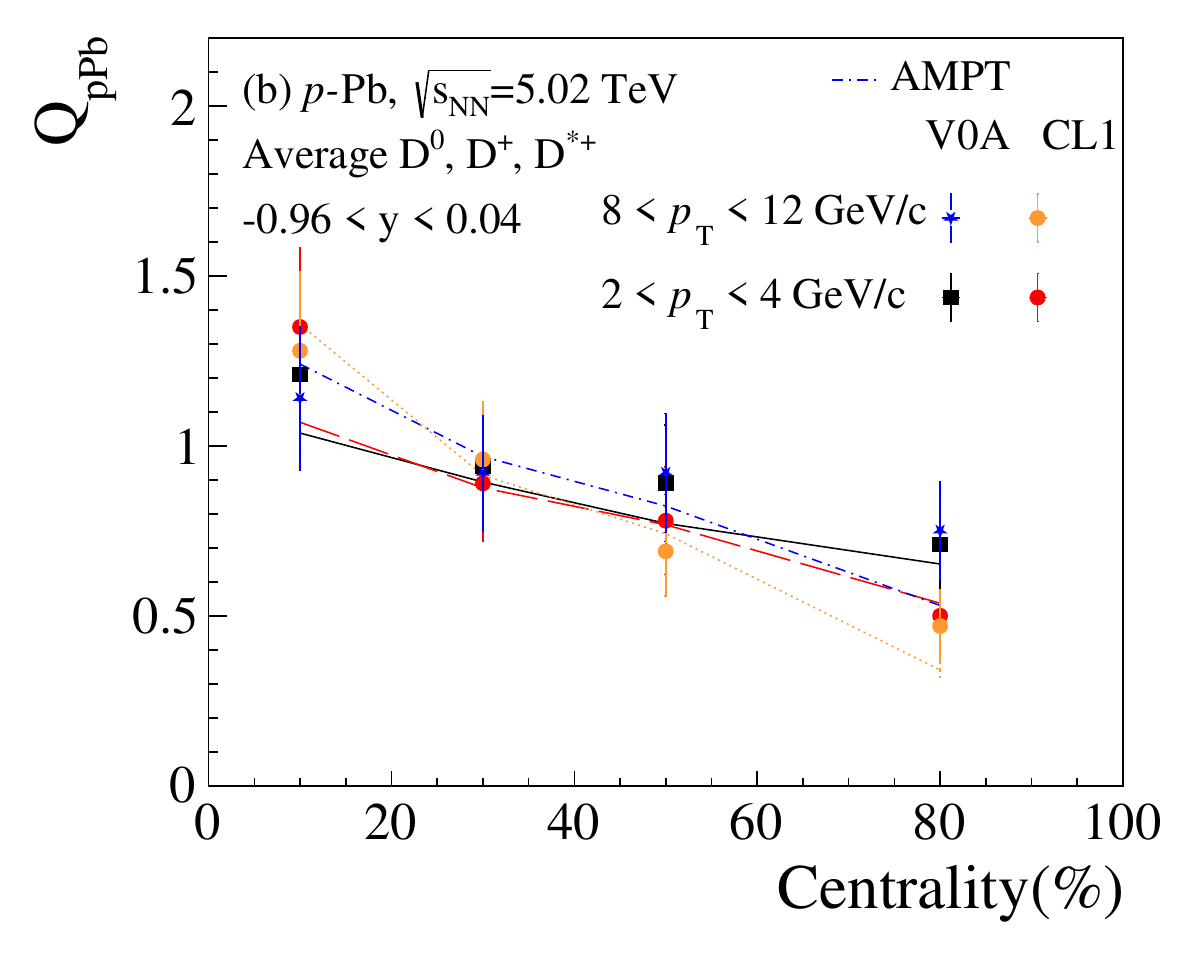}
\caption{The $\qppb$ of average D-mesons as function of the centrality from the CL1 and V0A event class from the AMPT model in comparison with the ALICE data~\cite{ALICE:2016cpm} in $p-$Pb collisions at $\snn=$ 5.02 TeV. 
}
\label{fig:8}
\end{figure}
%%%%%%%%%%%%%%%%%%%%%%%%%%%%%%%%%%%%%%%%%%%%%%%%%%%%%%%%%%%

We now use the improved AMPT model to study $D^0$ meson productions and $\rppb$ at different rapidities in $p-$Pb collisions at 5.02 TeV. In our recent work~\cite{Zhang:2022fum,Zhang:2023xkw}, the AMPT model is found to well describe the $D^{0}$ meson production and $\rppb$ at mid-rapidity. Here we further study these observables in different centralities and rapidities. Following the settings in Ref~\cite{Zhang:2022fum}, the Lund string fragmentation parameters are $a_{\rm{L}}=0.8$, $b_{\rm{L}}=0.7$ GeV$^{-2}$ for $p$+$p$ collisions while they are determined from the local nuclear scaling for $p-$Pb collisions~\cite{Zhang:2021vvp}.  Unless otherwise specified, the yield of $D^0$ meson represents the sum of the particles and the corresponding antiparticles, and the rapidity $y$ and $\eta$ are defined in the nucleon-nucleon centre-of-mass frame. 

The extracted value of $\delta$, along with its uncertainty determined from the $95\%$ confidence interval, as a function of rapidity is shown in Fig.\ref{fig:6}, where a smaller value of $\delta$ indicating a weaker Cronin is needed in the backward/forward rapidities. This observation aligns with measurements in $d$-Au collisions~\cite{PHENIX:2003qdw,STAR:2003pjh,PHOBOS:2003uzz}. Additionally, it is suggested that the influence of quantum evolution diminishes the Cronin enhancement at forward or backward rapidities for the asymmetric collisions~\cite{Kharzeev:2003wz}. 

\subsection{Centrality-dependent nuclear modification factor $\qppb$}

Here, we study the nuclear modification factors of averaged $D^0$, $D_{s}^+$ and $D^{*+}$ mesons as function of $\pt$ at mid-rapidity in four event class of $p-$Pb collisions at $\snn=$ 5.02 TeV. The nuclear modification factor is calculated as: 

\begin{equation}
\qppb=\frac{1}{\left\langle T_{\rm pPb}\right\rangle}\frac{d^2N(pA\to DX)/d\pt dy}{d^2\sigma(pp\to DX)/d\pt dy}.\label{eq7}
\end{equation}

Where $d^2N(pA\to DX)/d\pt dy$ is the yield of $D^0$ meson in a given centrality in $p-$Pb collisions, $d^2\sigma(pp\to DX)/d\pt dy$ is the production cross section of $D^0$ meson in $p$+$p$ collisions at the same energy, and $\langle T_{\rm pPb}\rangle$ is the average nuclear overlap function in a given centrality. Fig.~\ref{fig:7}(a) and (b) show the $\qppb$ using two different centrality estimators based on the multiplicity, CL1 and V0A, respectively.  Here, the centrality estimator CL1 uses charged particles within the rapidity region $|\eta|<1.4$, while V0A uses charged particles within $2.8<\eta<5.1$. As discussed in Ref.~\cite{ALICE:2016cpm}, $\qppb$ may be influenced by biases in centrality estimation unrelated to nuclear effects. Consequently, both sets of $\qppb$ data exhibit an ordering from low (60–100$\%$) to high (0–20$\%$) multiplicity. We observe that the AMPT model reproduces the trends seen in the experimental data for both CL1 and V0A event classes. Note that in the AMPT calculation, we incorporate $\langle T_{\rm pPb}\rangle$ values from experimental data when calculating $\qppb$. The AMPT model also replicates the behavior of smaller difference between the centrality classes in the V0A than that in the CL1. This is expected, as the rapidity gap between V0A and mid-rapidity D-meson cancelled out part of the event selection bias.

The averaged  $D^0$, $D_{s}^+$ and $D^{*+}$ mesons $\qppb$ obtained from CL1 and V0A esitimators for $2<\pt<4$ GeV/$c$ and $8<\pt<12$ GeV/$c$ as function of centrality are shown in Fig.~\ref{fig:8}. We observe that the AMPT model is consistent with the experiment data for the both low and high $\pt$ regions at CL1 and V0A within the uncertainty. However, it slightly under-esimates the data at 0-20$\%$ centrality for $2<\pt<4$ GeV/$c$.

\subsection{$\pt$-differential cross sections}

\begin{figure}[!htb]
\includegraphics[scale=0.43]{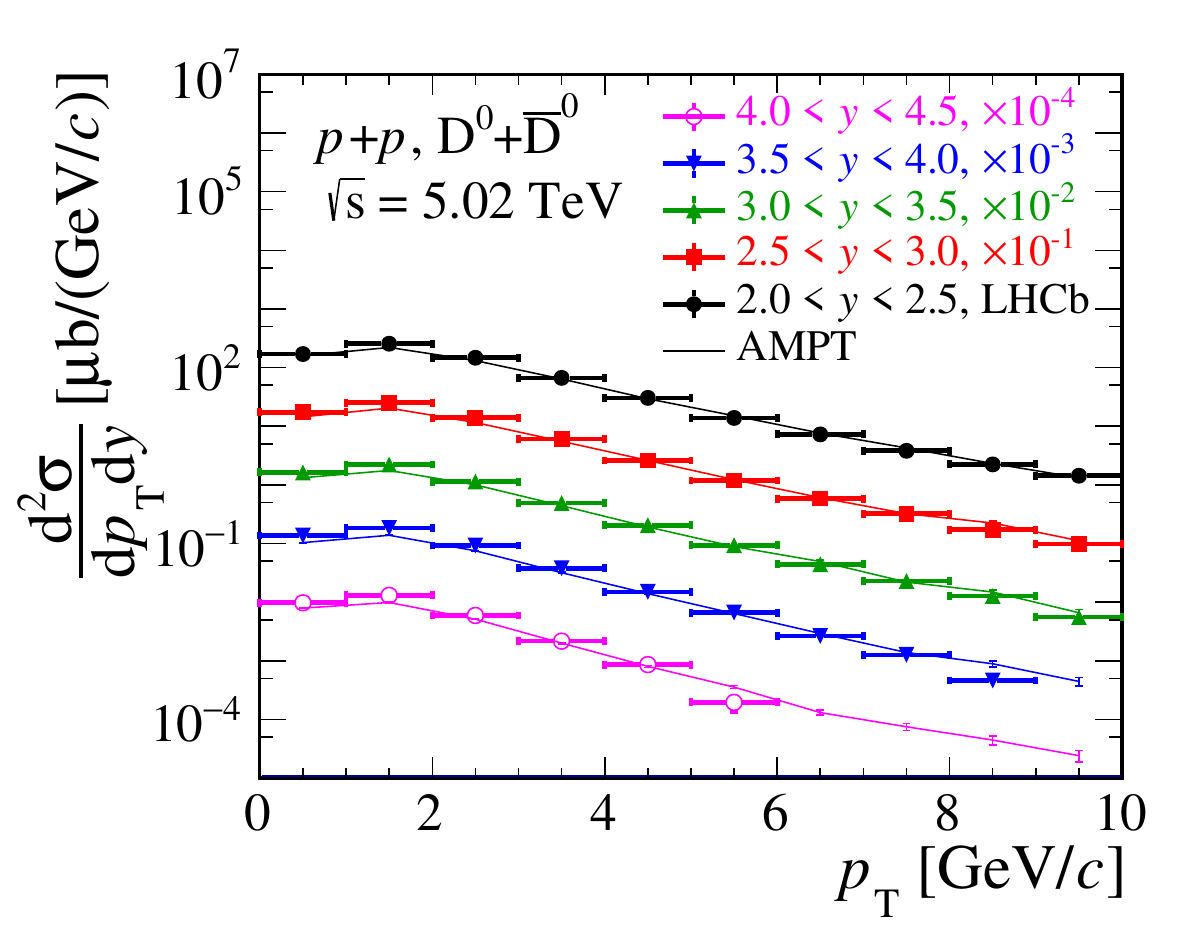}
\caption{ The double differential production cross section of $D^0+\bar{D^0}$ meson as function of $\pt$ in $p$+$p$ collisions at $\sqrt{s}=$ 5.02TeV in forward rapidities from AMPT model in comparison with LHCb data~\cite{LHCb:2016ikn}.
}
\label{fig:9}
\end{figure}

We next present the double differential production cross sections for $D^0$ in $p$+$p$ collisions at $\sqrt{s}=$ 5.02 TeV from AMPT model. In Fig.~\ref{fig:9} the results in different rapidities are compared with the LHCb data~\cite{LHCb:2016ikn}. The AMPT model in general provides an good description of the data for all rapidity intervals.
%%%%%%%%%%%%%%%%%%%%%%%%%%%%%%%%%%%%%%%%%%%%%%%%%%%%%%%%%%%
%%%%%%%%%%%%%%%%%%%%%%%%%%%%%%%%%%%%%%%%%%%%%%%%%%%%%%%%%%%

%%%%%%%%%%%%%%%%%%%%%%%%%%%%%%%%%%%%%%%%%%%%%%%%%%%%%%%%%%%
%\begin{figure}[!htb]
%\centering
%\begin{tikzpicture}
%\includegraphics[scale=0.45]{fig5.pdf}
%\caption{The nuclear modification factors $R_{pA}$ of different hadron species from AMPT model in comparison with data in Pb+p collisions.
%}
%\label{fig:5}
%\end{figure}
\begin{figure*}
    \begin{minipage}[t]{0.5\linewidth}
    \centering
    \includegraphics[scale=0.43]{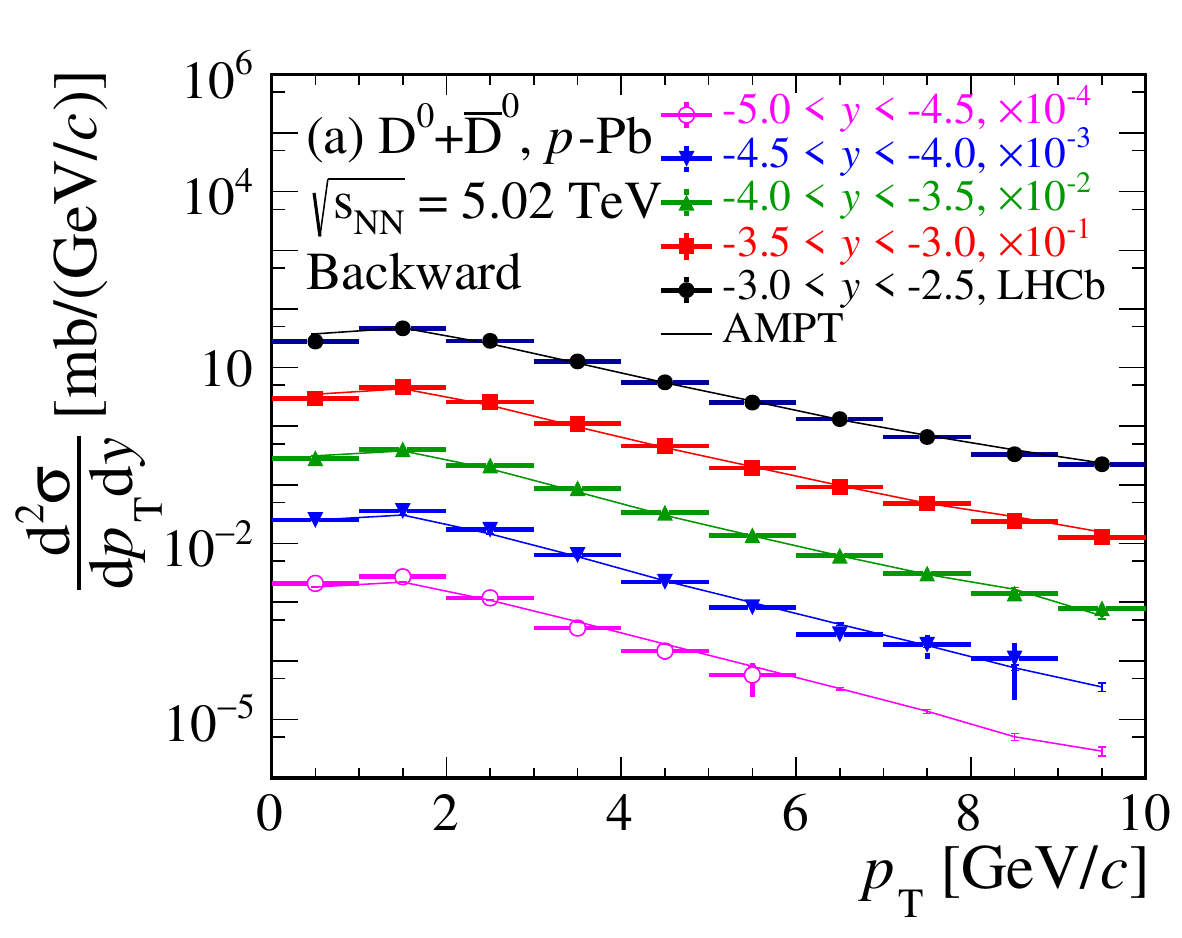}    
  \end{minipage}%
  \begin{minipage}[t]{0.5\linewidth}
    \centering
    \includegraphics[scale=0.43]{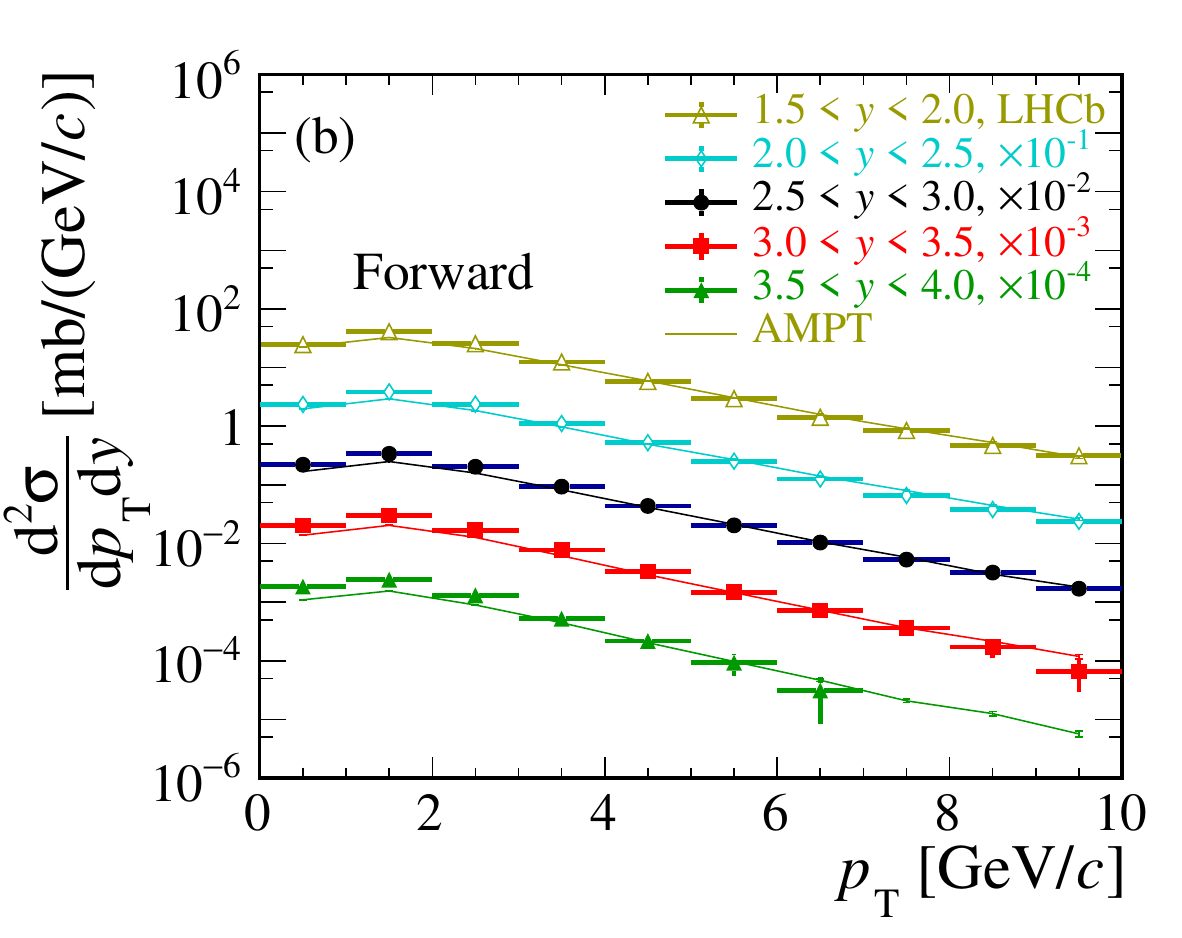}
     \end{minipage}
    \caption{The double differential production cross section of $D^0+\bar{D^0}$ meson in the (a): backward and (b): forward rapidities as function of $\pt$ in minimum-bias $p-$Pb collisions at $\snn=$ 5.02 TeV from the AMPT model in comparison with LHCb data~\cite{LHCb:2017yua}.}
     \label{fig:10}
  \end{figure*}
%%%%%%%%%%%%%%%%%%%%%%%%%%%%%%%%%%%%%%%%%%%%%%%%%%%%%%%%%%%

Figure~\ref{fig:10}(a) and (b) show the cross sections of $D^0$ production calculated by the AMPT model for $p$–Pb collisions at 5.02 TeV in the backward and forward rapidity ranges, respectively. The model results are compared with experimental data at LHCb. Within the rapidity intervals $-5<y<-2.5$, the AMPT model well reproduces both the magnitude and trends of the $D^0$ production cross section. However, on the proton-going side, the AMPT model slightly underestimates the $D^0$ production, particularly in $\pt\leq$2 GeV/$c$ in the rapidity intervals $3.0<y<4.0$, exhibiting the same behavior as observed in $p$+$p$ collisions. Note that a value of $\delta$ smaller than 5.0 in the Eq.~\ref{eq2} is needed in order to describe the transverse momentum dependence in the backward rapidities. Therefore, the $\delta$ value is assigned according to the central value of Fig.~\ref{fig:2} based on the rapidity of the initial $c\bar{c}$ pairs in this calculations.

\subsection{$\pt$ and $y$ dependence of nuclear modification factor $\rppb$}

%In Fig.~\ref{fig:8} we analyze the ratio of $D^{+}/D^{0}$, $D^{+}_{s}/D^{0}$ and $D^{*+}/D^{0}$  in $p$+$p$ and minimum bias $p-$Pb collisions at 5.02 TeV around mid-rapidity, these results are then compared with the experimental data. We see that the improved AMPT model provide good descriptions of all three results in both collision systems. Particularly noteworthy is its ability to capture the rising trend of the $D^{*+}/D^{0}$ ratio as a function of $\pt$. In our recent study~\cite{Zhang:2022fum}, the improved AMPT model has been shown to effectively reproduce the $\pt$ spectrum of $D^{0}$ mesons at mid-rapidity in $p$+$p$ and minimum bias $p-$Pb collisions. Hence, the AMPT model effectively describes the production of $D^0$, $D^{+}$, $D^{*+}$, and $D_{s}^{+}$ mesons as well as their ratios in both collision systems.
%It should be noted that we made a correction to the branching ratio of $D^{+}$ decay in the HIJING initial condition model. Specifically, we replaced the value of 0.49 for $D^{+}\to D^0$ with the updated value of 0.68.
%%%%%%%%%%%%%%%%%%%%%%%%%%%%%%%%%%%%%%%%%%%%%%%%%%%%%%%%%%%

\begin{figure*}
  \begin{minipage}[t]{0.5\linewidth}
    \centering
    \includegraphics[scale=0.43]{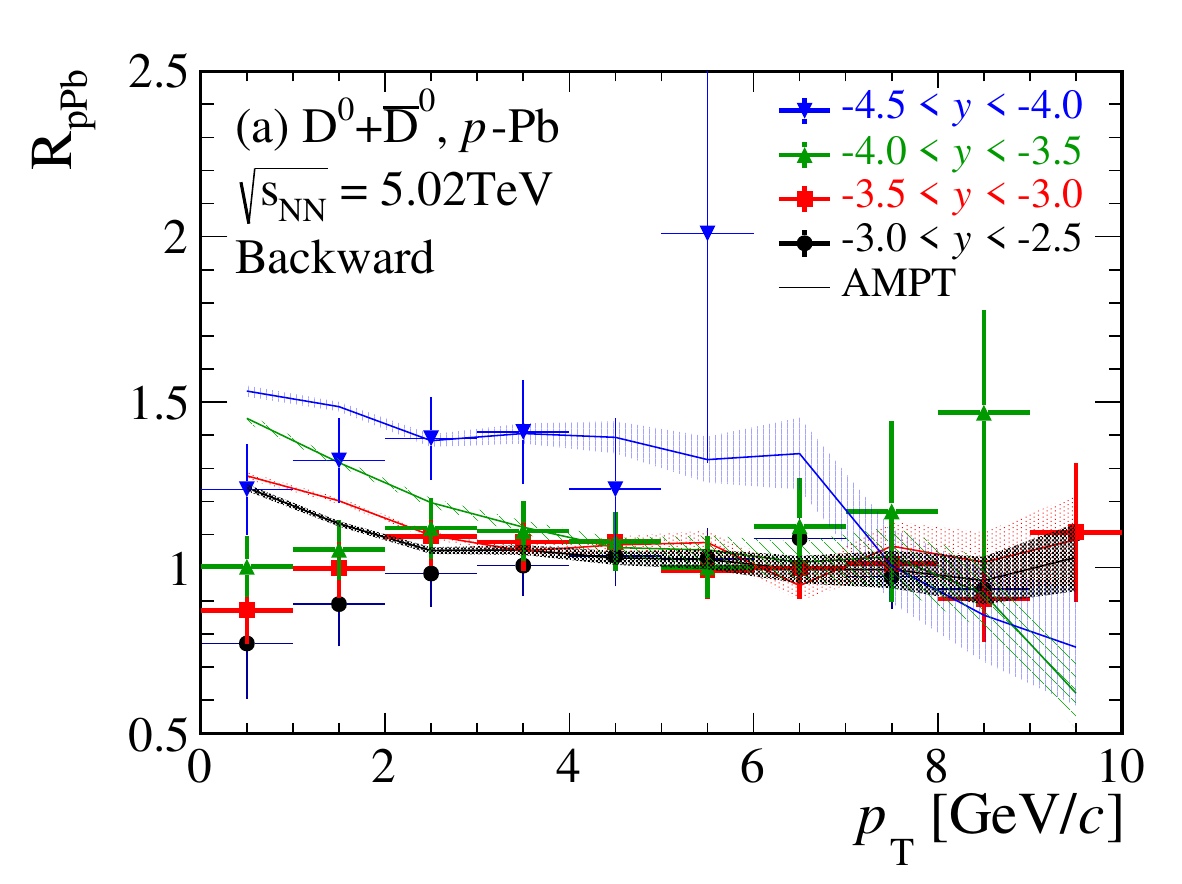}
   %   \label{fig:side:a}
  \end{minipage}%
  \begin{minipage}[t]{0.5\linewidth}
    \centering
    \includegraphics[scale=0.43]{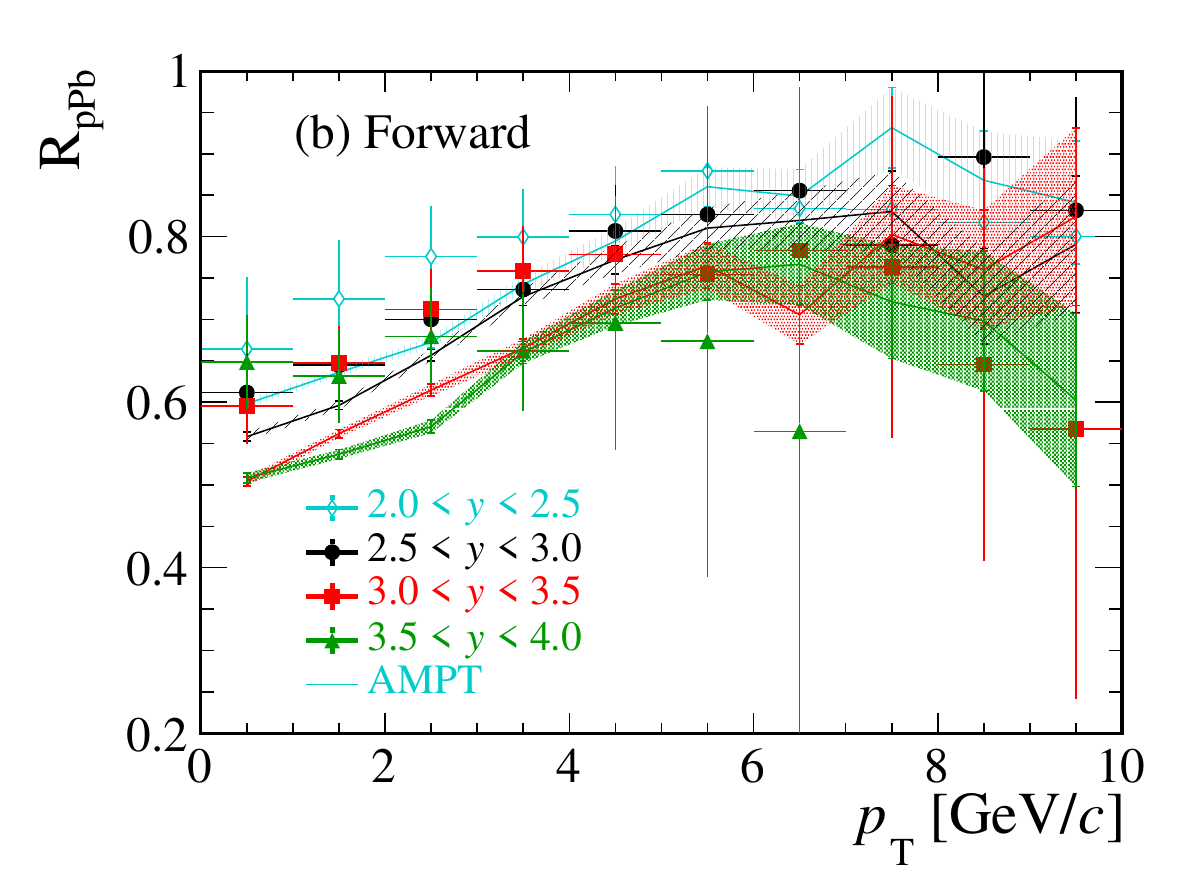}
   %  \label{fig:side:b}
  \end{minipage}
  \caption{ The $\rppb$ of $D^{0}$ mesons (a): at $-4.5<y<-2.5$, (b): at $2.0<y<4.0$, with a rapidity bin width of 0.5, as a function of $\pt$ in minimum bias $p-$Pb collisions at $\snn=$ 5.02 TeV.The results, obtained from the AMPT model with $\delta$ values derived from $\chi^2$ fitting, are compared with LHCb data (symbols)~\cite{LHCb:2017yua}.
}
 \label{fig:11}
\end{figure*}
%%%%%%%%%%%%%%%%%%%%%%%%%%%%%%%%%%%%%%%%%%%%%%%%%%%%%%%%%%%
We then investigate the multiplicity integrated nuclear modification factors $\rppb$ of  $D^0$ meson as function of $\pt$ and $y$. The $\rppb$, which is the ratio of double differential production cross section of $D^0$meson in $p-$Pb collisions to that in $p$+$p$ collisions at the same rapidity and collision energy, is calculated by:

\begin{equation}
\rppb=\frac{1}{A}\frac{d^2\sigma(pA\to DX)/d\pt dy}{d^2\sigma(pp\to DX)/d\pt dy},\label{eq8}
\end{equation}
where $A=208$ is the atomic number of the lead nucleus. The $\rppb$ is often used to quantify the modifications of the nuclear interactions with respect to binary nucleon nucleon collisions. It was found that the $\rppb$ is sensitive to both cold nuclear matter effects, such as the Cronin broadening and nuclear shadowing, and hot nuclear effects, such as jet quenching and energy loss.

For the $\rppb$ of $D^0$ meson in $(-4.5<y<-2.5)$ and $(2.0<y<4.0)$ rapidities with a rapidity bin width of 0.5, the respective results as function of $\pt$ together with the experimental data are shown in Fig.~\ref{fig:11}(a) and  ~\ref{fig:11}(b). The $\rppb$ values in the experimental data generally evolve from lower to higher values from the $3.5<y<4.0$ to the $-4.5<y<-4.0$ rapidity, particularly in the low $\pt$ regions ($\pt\leq$2.5 GeV/$c$).  We observe that the AMPT model captures these trends in general. In addition, the AMPT model reasonably reproduces the $\rppb$ data for all rapidities with $\pt\geq$ 2 GeV/$c$. However, it overpredicts the $\rppb$ at $\pt<$ 2.5GeV/$c$ in $-4.5<y<-2.5$, but underpredicts it in $2.0<y<4.0$.
% shows the $\rppb$ of $D^{+}$, $D_{s}^{+}$ and $D^{*+}$ in the minimum bias $p-$Pb collisions around mid-rapidity. The lines represent the results from AMPT model while the symbols are experimental data. Apparently, the experimental measurements for the nuclear modification factor of all D mesons tend to unity, while the predictions from the models are quite identical and describe the corresponding data within the uncertainties.

\begin{figure*}
  \begin{minipage}[t]{0.5\linewidth}
    \centering
    \includegraphics[scale=0.43]{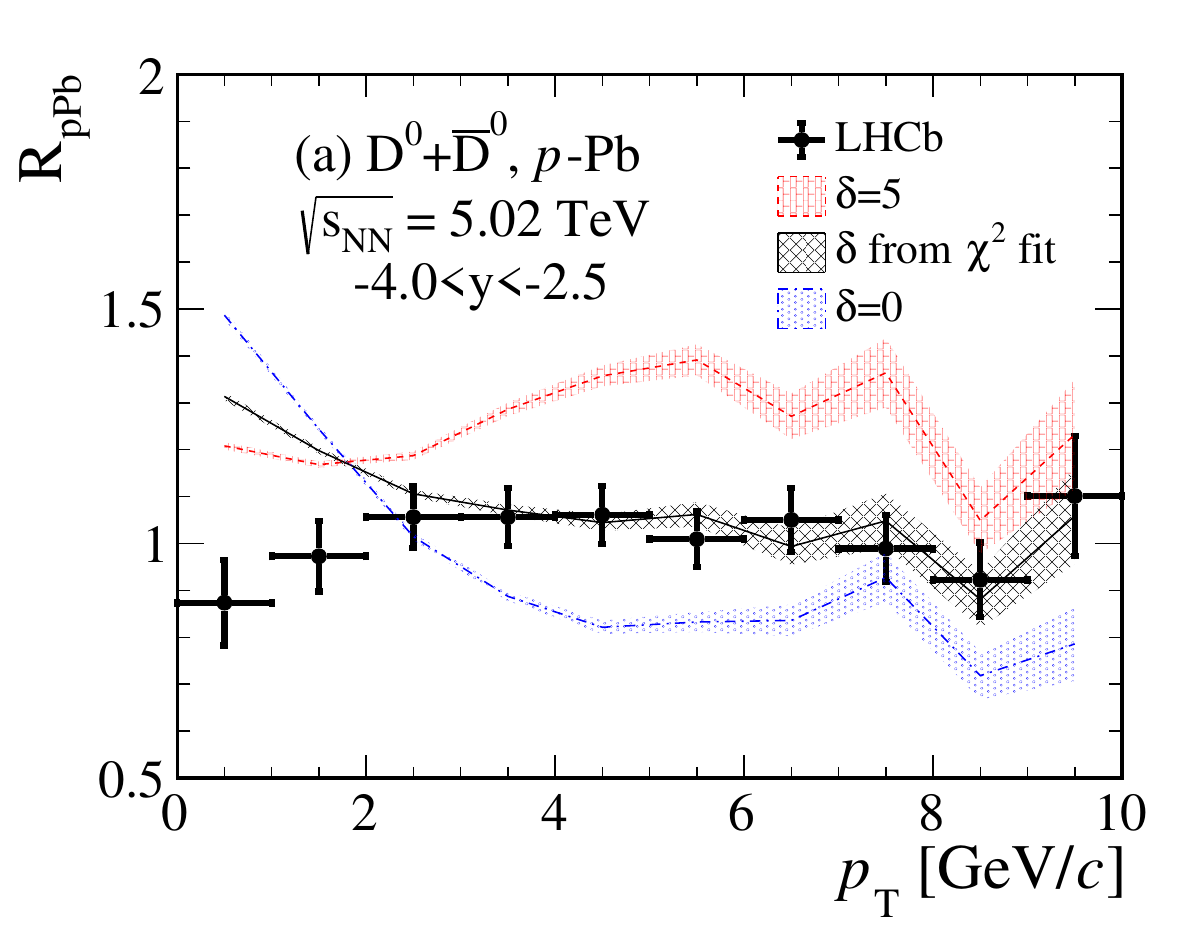}
   %   \label{fig:side:a}
  \end{minipage}%
  \begin{minipage}[t]{0.5\linewidth}
    \centering
    \includegraphics[scale=0.43]{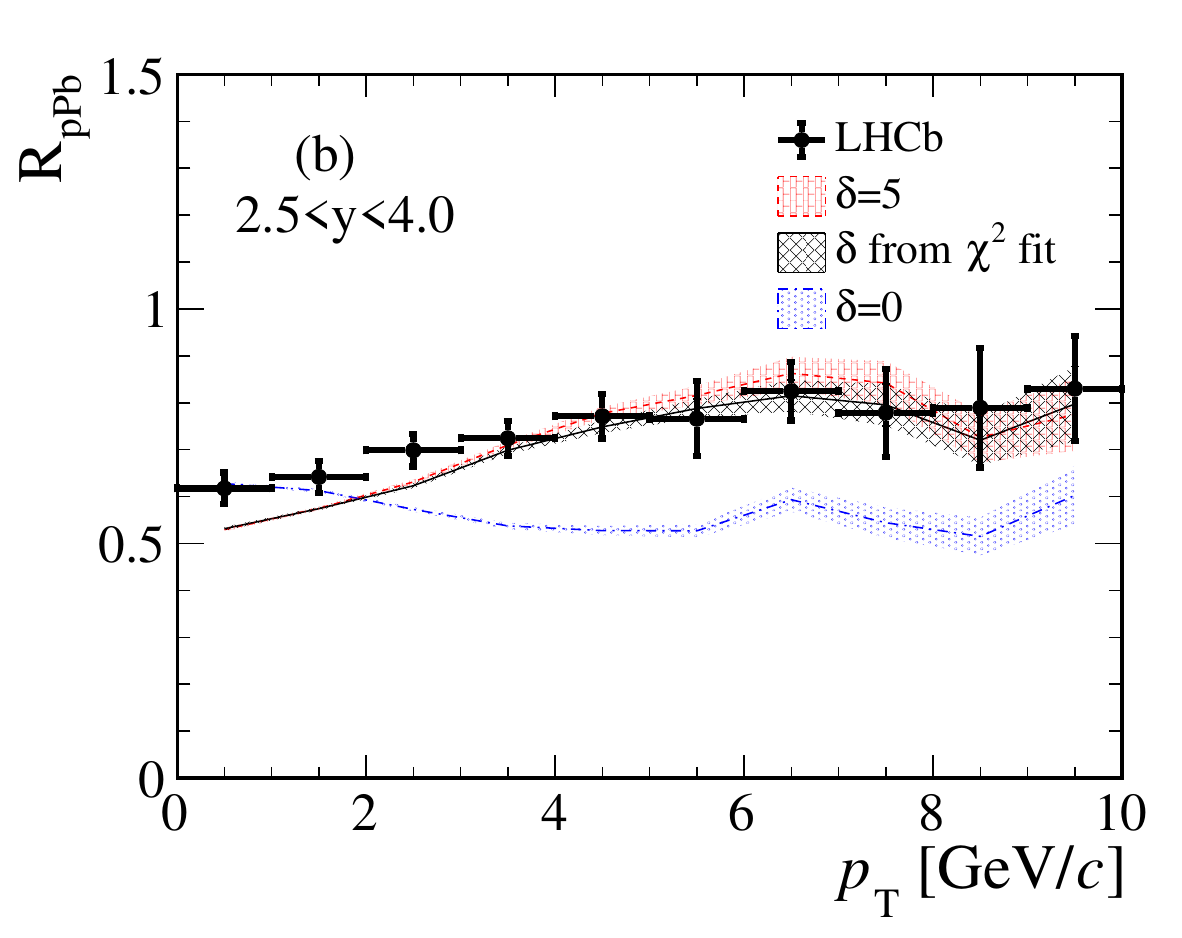}
   %  \label{fig:side:b}
  \end{minipage}
  \caption{The $\rppb$ of $D^{0}$ meson (a): within$(-4.0<y<-2.5)$ and (b): within $(2.5<y<4.0)$ as function of $\pt$ in minimum bias $p-$Pb collisions at $\snn=$ 5.02 TeV from the AMPT model with $\delta=0$ (dot-dashed), $\delta$ from the fit (solid) and 5 (dashed) in comparison with LHCb data (symbols)~\cite{LHCb:2017yua}.
}
 \label{fig:12}
\end{figure*}

Figure~\ref{fig:12} shows our results of the $\rppb$ for $D^0$ meson at $2.5<|y|<4$ in comparison with the LHCb data at 5.02 TeV. The experimental measurement of the nuclear modification factor is close to unity at backward rapidity ($\rppb \approx1$), whereas a strong suppression trend is shown in the forward region. This behavior can be partially attributed to the nuclear shadowing effect, as the suppression from shadowing in the forward rapidity corresponds to the small-$x$ region, while the anti-shadowing effect in the backward rapidity corresponds to the middle $x$~\cite{LHCb:2017yua}.
We observe a similar behavior in the AMPT model, although it overpredicts the $\rppb$ at $\pt<$ 2 GeV/$c$ in backward rapidities. The effects of $\delta$ are studied, and it is observed that the effect is more significant in the backward rapidity. As depicted, the $\rppb$ with $\delta = 0$ are more suppressed, especially at high $\pt$, for both rapidity ranges. While $\delta=5$ can effectively reproduce the $\rppb$ in the forward rapidity, it significantly overpredicts the data in the backward region. We observe that the $\delta$ values derived from the $\chi^2$ fit can in general describe the data. It is understandable that a ``weak" Cronin enhancement is needed for the $D^0$ meson in the forward rapidity compared to that in the mid-rapidity, as no significant Cronin enhancement is observed in the forward rapidity compared to that in mid-rapidity in $d$-Au collisions at RHIC~\cite{PHENIX:2003qdw,STAR:2003pjh,PHOBOS:2003uzz}. The reason for a ``weak" Cronin effect being observed in the forward rapidities remains an open question and worth further studies.
%This observation in our work is likely attributed to the less frequent nucleon-nucleon interactions in this region.
%%%%%%%%%%%%%%%%%%%%%%%%%%%%%%%%%%%%%%%%%%%%%%%%%%%%%%%%%%%
\begin{figure}[!htb]
\includegraphics[scale=0.43]{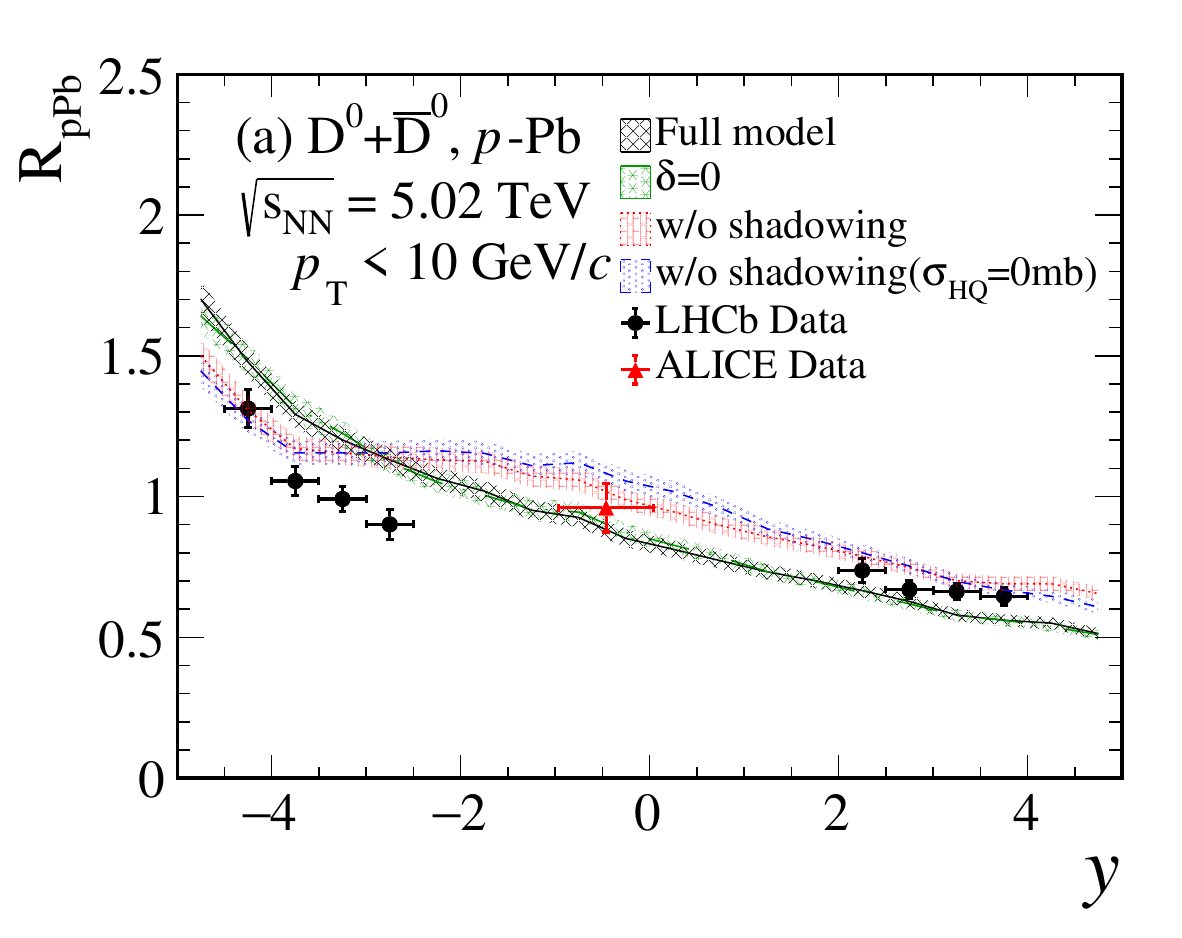}
\caption{ The $\rppb$ of $D^{0}$ mesons as a function of rapidity, integrated over $\pt<$10 GeV/$c$, in minimum bias $p-$Pb collisions at $\snn=$5.02 TeV. The results are shown from the AMPT model (solid line), the model with $\delta=0$(dot dashed line), the model without the shadowing effect (dotted line), and the model without both shadowing and parton cascade effects (dashed line), alongside with experimental data (symbols)~\cite{LHCb:2017yua}.
}
\label{fig:13}
\end{figure}
%%%%%%%%%%%%%%%%%%%%%%%%%%%%%%%%%%%%%%%%%%%%%%%%%%%%%%%%%%%

In what follows we present the results of $\rppb$ as function of rapidity for $D^{0}$ meson integrated over $0<\pt<10$GeV/$c$ in the minimum bias $p-$Pb collisions at $\snn=$ 5.02 TeV. The calculation of $\rppb$ follows Eq.~\ref{eq8}. The results obtained by the AMPT model are compared with the available experimental data from LHCb~\cite{LHCb:2017yua} and ALICE collaborations~\cite{ALICE:2019fhe}, as displayed in Fig.~\ref{fig:13}. The AMPT model adequately explains the $\rppb$ at mid and forward rapidities. However, it overestimates the $\rppb$ in the backward rapidities. The Cronin effect has been studied, and in general, we observe no significant effect resulting from the $\pt$ integration of the cross sections when calculating the $\rppb$.
We further studied the impact of nuclear shadowing by disabling it in the $p-$Pb collisions in the AMPT calculations. It is observed that the nuclear shadowing suppresses $D^0$ $\rppb$ in the mid- and forward rapidity; however, the effects in the backward range are the opposite. It is intriguing to observe the rapidity-dependent nuclear shadowing effect on the $\rppb$ of $D^{0}$ meson, which depends on the variation of momentum fraction $x$ across different rapidity regions. The effect of parton cascade is then studied by turning off the parton scattering of the heavy quarks ($\sighq$=0mb) in the AMPT calculations. The presence of heavy quark scattering suppresses $D^{0}$ meson $\rppb$ at mid-rapidity while enhancing it in the far forward and backward regions. As a result, the $\rppb$ from the AMPT model without nuclear shadowing becomes more flat. Our observations imply that the $\rppb$ shape in experiments arises from a combination of the Cronin effect, nuclear shadowing, and interactions within the de-confined parton matter of the system.

\subsection{Production cross sections}

\begin{figure}[!htb]
\includegraphics[scale=0.43]{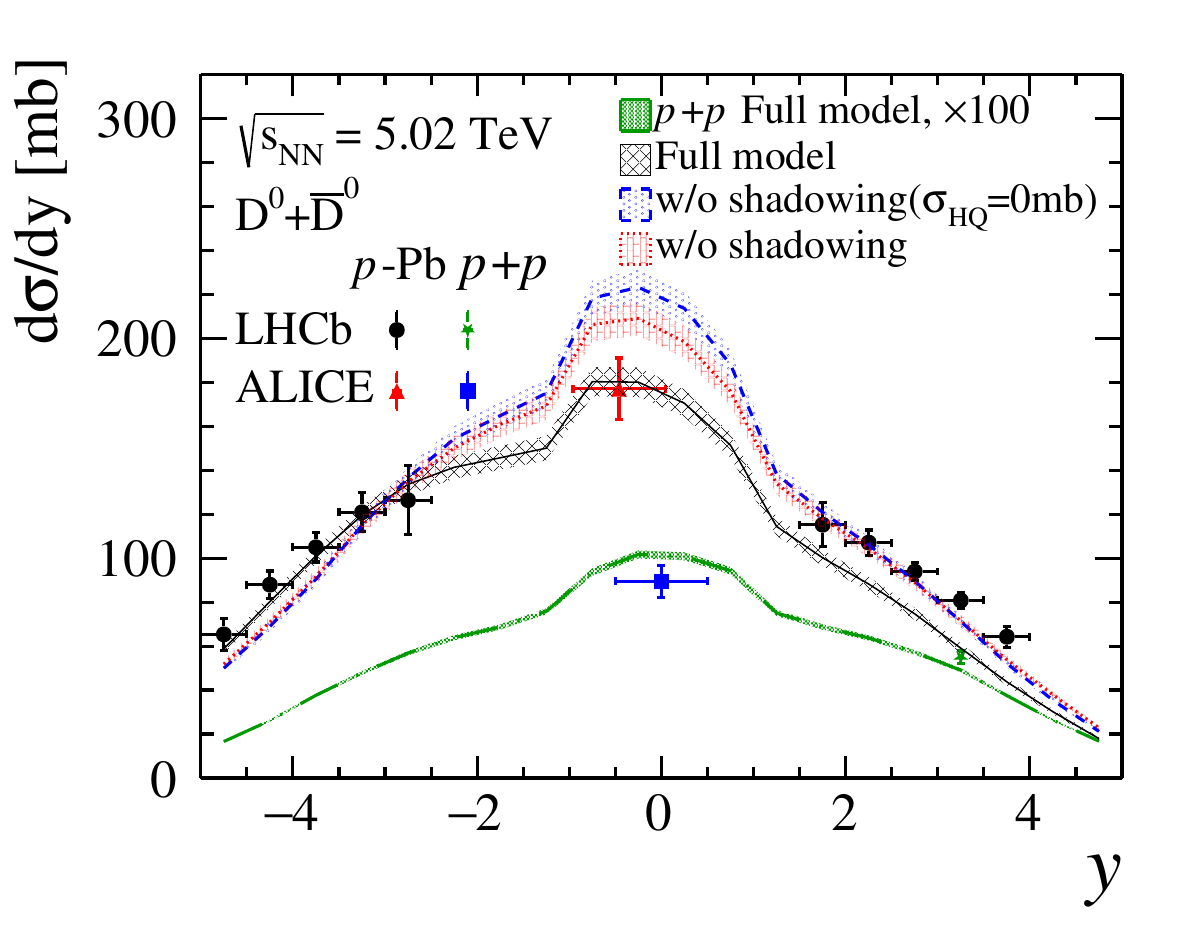}
\caption{ The $D^{0}$ mesons production cross section as a function of rapidity, integrated over $\pt<$10 GeV/$c$, in both $p$+$p$ and minimum bias $p-$Pb collisions at $\snn=$5.02 TeV. The results are shown from the AMPT model in $p$+$p$ collsions (dot dashed line), and the AMPT model in $p-$Pb collisions (solid line), the model without the shadowing effect (dotted line), and the model without both shadowing and parton cascade effects (dashed line), alongside with experimental data (symbols)~\cite{LHCb:2017yua,ALICE:2019nxm,LHCb:2016ikn}.
}
\label{fig:14}
\end{figure}

Finally, we investigate the production cross section of $D^{0}$ mesons as a function of rapidity, integrated over $0<\pt<10$ GeV/$c$, in $p$+$p$ and minimum bias $p-$Pb collisions at 5.02 TeV. The results obtained from the AMPT model are compared with experimental data, as illustrated in Fig.~\ref{fig:14}. For $p$+$p$ collisions, the AMPT model slightly overestimates $D^0$ production at mid-rapidity and underpredicts it in the forward region.
Similarly, the AMPT model well describes the $D^0$ production cross section in the $p-$Pb collisions at mid- and backward rapidities, but underestimates it by approximately $20\%$ in the positive range. Furthermore, we examine the impact of nuclear shadowing and parton scattering by disabling them in the $p-$Pb collisions in the AMPT calculations. We observe that nuclear shadowing significantly suppresses $D^0$ production in the mid- and forward rapidity regions, while its effects in the backward range are only marginal. Hence, the under-prediction of the production cross section at forward rapidities in $p-$Pb collisions could be attributed to the descrepency in $p$+$p$ collisions and nuclear shadowing. It's worth noting that there remains a significant uncertainty regarding nuclear shadowing of gluons~\cite{Helenius:2012wd}, which we have not explored in this study. Additionally, we investigate the effect of the parton cascade by turning off parton scattering of the heavy quarks($\sighq$=0 mb) in the AMPT calculations. Similar to the $\rppb$ results, the presence of parton scattering suppresses $D^{0}$ meson production at mid-rapidity while enhancing it in the far forward and backward regions.
%%%%%%%%%%%%%%%%%%%%%%%%%%%%%%%%%%%%%%%%%%%%%%%%%%%%%%%%%%%
%%%%%%%%%%%%%%%%%%%%%%%%%%%%%%%%%%%%%%%%%%%%%%%%%%%%%%%%%%%

\section{Discussion}
\label{sec:dis}

In the present AMPT model, interactions between the heavy quarks and the QGP medium are modeled by two-body elastic scatterings using ZPC, while heavy quark radiative energy loss has not been considered. Previous studies suggest that the heavy quark collisional energy loss depends linearly on the path length, $L$, while the radiative energy loss is proportional to $L^{2}$.
Therefore, for small systems such as $p-$Pb collisions, the neglection of the radiative energy loss in the AMPT model is expected to have only a small effect on heavy quark observables.
In the meanwhile, the implementation of the radiative energy loss mechanism of heavy quark in the ZPC model would be worthwhile since it would enable the AMPT model to better describe the heavy flavor physics in larger systems. Therefore, future study of heavy flavor $\raa$ and $v_2$ in heavy-ion collisions is planned.

In a previous study, the Cronin effect has been proposed to be the key component in resolving the $\rppb$ and $v_2$ puzzles in $p-$Pb collisions at LHC. In this work, we further examine the importance of including the Cronin effect in describing the $\pt$ spectra and $\rppb$ of $D^0$ meson in the forward and backward rapidities. 
Our results indicate a non-trivial rapidity dependence of the strength of the Cronin effect in $p-$Pb collisions.
In addition, the effect of the Cronin is expected to magnify with increased collision system size. Therefore, further higher precision experimental measurements at different rapidities and the consideration of the Cronin effect in studies of large systems will be necessary to advance the understanding of the Cronin effect.

\section{Summary }
\label{sec:summary}
Using a multi-phase transport model, we have investigated the $\pt$ spectra, multiplicity dependent nuclear modification factor ($\qppb$) and multiplicity integrated $\rppb$ of $D$ mesons in different rapidity intervals in minimum bias $p-$Pb collisions at 5.02 TeV. The AMPT model has been improved by extracting the heavy quarks from HIJING initial condition as well as incorporating the transverse momentum broadening (i.e., the Cronin effect) and independent fragmentation for charm quarks.  Our results demonstrate that the improved model can reasonably describe $D^0$ meson spectra and corresponding $\qppb$ up to 10 GeV/$c$. Furthermore, the AMPT model captures the general rapidity dependence of $D^0$ meson production and $\rppb$. We observe that a rapidity-dependent strength of the Cronin effect is necessary for describing the $\rppb$ data at different rapidities. This study contributes to a better understanding of cold-nuclear-matter effects in relativistic heavy-ion collisions.

\begin{acknowledgements}
This work is supported in part by the National Key Research and Development Program of China under Contract No. 2022YFA1604900 (S.S); the National Natural Science Foundation of China (NSFC) under contract No. 12175084 (S.S), the Fundamental Research Funds for the Central Universities, China University of Geosciences  (Wuhan) with No. G1323523064 (L.Z), and the National Science Foundation under Grant No. 2310021 (Z.-W.L.).

%If you'd like to thank anyone, place your comments here
%and remove the percent signs.
\end{acknowledgements}

% BibTeX users please use one of
%\bibliographystyle{spbasic}      % basic style, author-year citations
%\bibliographystyle{spmpsci}      % mathematics and physical sciences
%\bibliographystyle{spphys}       % APS-like style for physics
%\bibliography{}   % name your BibTeX data base

% Non-BibTeX users please use

\end{document}